\documentclass[10pt,journal]{IEEEtran}

\usepackage{tikz}
\usepackage{booktabs} 
\usepackage{tikz}
\usetikzlibrary{topaths,calc}
\newcommand{\tabincell}[2]{\begin{tabular}{@{}#1@{}}#2\end{tabular}}

\usepackage{caption}
\usepackage[english]{babel} 
\usepackage{amsthm}
\usepackage{array}
\usepackage{mdwmath}
\usepackage{mdwtab}
\usepackage{latexsym,bm,amsmath,amssymb}
\usepackage{bm}
\usepackage{subfigure}
\usepackage{comment}
\usepackage{wrapfig}
\usepackage{threeparttable,booktabs}
\usepackage{footnote}
\usepackage{mathrsfs}
\usepackage{algorithm}
\usepackage{algorithmic}
\usepackage{enumitem}
\newlist{steps}{enumerate}{1}
\setlist[steps, 1]{label = \textbf{Step \arabic*:}}
\usepackage{xcolor}

\DeclareMathAlphabet{\pazocal}{OMS}{zplm}{m}{n}

\newcommand{\Yb}{\pazocal{Y}}
\newcommand{\Tb}{\pazocal{T}}
\newcommand{\Zb}{\pazocal{Z}}

\newcommand{\Cross}{\mathbin{\tikz [x=1ex,y=1.6ex,line width=.13ex] \draw (0,0) -- (0.85,0.85) (0,0.85) -- (0.85,0);}}%

\usepackage{footnote}
\makesavenoteenv{tabular}
\makesavenoteenv{table}
\newcommand{\squeezeup}{\vspace{-3mm}}

\makeatletter

\def\ps@IEEEtitlepagestyle{%
\def\@oddhead{\parbox[t][\height][t]{\textwidth}{\centering
This work has been accepted for publication in IEEE Transactions on Information Forensics and Security.
}\hfil\hbox{}}%
\def\@evenhead{\scriptsize\thepage \hfil \leftmark\mbox{}}%
}

\makeatother
\begin{document}
%

\title{Machine Learning-Based Delay-Aware UAV Detection and Operation Mode Identification over Encrypted Wi-Fi Traffic}

\author{Amir~Alipour-Fanid,~\IEEEmembership{Student Member,~IEEE,}
		Monireh~Dabaghchian,~\IEEEmembership{Member,~IEEE,}
		Ning~Wang,~\IEEEmembership{Member,~IEEE,}
		Pu~Wang,~\IEEEmembership{Student Member,~IEEE,}
		Liang~Zhao,~\IEEEmembership{Member,~IEEE,}
        Kai Zeng,~\IEEEmembership{Member,~IEEE,}
\thanks{Manuscript received; revised; accepted.
Date of publication; date of current version.
This work was partially supported by the National Security Agency (NSA) under grants No. H98230-16-1-0356, No.H98230-18-1-0343, Virginia’s Commonwealth Cyber Initiative (CCI), National Science Foundation (NSF) under grants No. 1755850, No. 1841520, No. 1907805, Jeffress Trust Award, and NVIDIA GPU Grant.
This paper is an extended version of the workshop paper presented in IEEE CNS International Workshop On
Cyber-Physical Systems Security (CPS-Sec) Washington DC, USA, June 2019.
}

\thanks {Amir Alipour-Fanid, Ning Wang, Pu Wang and Kai Zeng are with the Department of Electrical and Computer Engineering, George Mason University, Fairfax, VA, 22030 USA (e-mail: \{aalipour, nwang5, pwang20, kzeng2\}@gmu.edu).}

\thanks{Monireh Dabaghchian is with the Department of Computer Science, Morgan State University, Baltimore, MD, 21251 USA (e-mail: monireh.dabaghchian@morgan.edu).}

\thanks{Liang Zhao is with the Department of Information Science and Technology, George Mason University, Fairfax, VA, 22030 USA (e-mail: lzhao9@gmu.edu).}
}

\maketitle

\begin{abstract}
The consumer unmanned aerial vehicle (UAV) market has grown significantly over the past few years.
Despite its huge potential in spurring economic growth by supporting various applications, the increase of consumer UAVs poses potential risks to public security and personal privacy.
To minimize the risks, efficiently detecting and identifying invading UAVs is in urgent need for both invasion detection and forensics purposes.
Aiming to complement the existing physical detection mechanisms, we propose a machine learning-based framework for fast UAV identification over encrypted Wi-Fi traffic.
It is motivated by the observation that many consumer UAVs use Wi-Fi links for control and video streaming.
The proposed framework extracts features derived only from packet size and inter-arrival time of encrypted Wi-Fi traffic, and can efficiently detect UAVs and identify their operation modes.
In order to reduce the online identification time, our framework adopts a re-weighted $\ell_1$-norm regularization, which considers the number of samples and computation cost of different features.
This framework jointly optimizes feature selection and prediction performance in a unified objective function.
To tackle the packet inter-arrival time uncertainty when optimizing the trade-off between the detection accuracy and delay, we utilize maximum likelihood estimation (MLE) method to estimate the packet inter-arrival time.  
We collect a large number of real-world Wi-Fi data traffic of eight types of consumer UAVs and conduct extensive evaluation on the performance of our proposed method.
Evaluation results show that our proposed method can detect and identify tested UAVs within 0.15-0.35s with high accuracy of 85.7-95.2\%.
The UAV detection range is within the physical sensing range of 70m and 40m in the line-of-sight (LoS) and non-line-of-sight (NLoS) scenarios, respectively.
The operation mode of UAVs can be identified with high accuracy of 88.5-98.2\%.

\end{abstract}

{\bf \emph{Index Terms -} Unmanned aerial vehicle (UAV) detection, machine learning, encrypted Wi-Fi traffic classification.}

\IEEEpeerreviewmaketitle

\section{Introduction}
\label{Introduction}

\IEEEPARstart{I}{n} the past few years, we have seen a significant growth of the consumer unmanned aerial vehicle (UAV) market for personal recreation.
Despite its huge potential in spurring economic growth, the significant increase of consumer UAVs raises lots of issues regarding airspace management, public security, and personal privacy \cite{ARYA2017}.
It was reported that an Army chopper was struck by an illegally flying drone over a residential neighborhood in September 2017 \cite{RT2017crash}.
In April 2016, a UAV  was  peeping outside a teenager's bedroom window in Massachusetts \cite{RT2016peeping}.
In January 2015, a small UAV crashed on the White House lawn bringing the worry about  security measures \cite{shear2015white}.

To deal with these threats, consumer UAV registration mechanisms, started by Federal Aviation Administration (FAA), have been promoted worldwide, which can help law enforcement officials to handle the UAV and its owner information \cite{FAA}.
UAV-restricted zone and geo-fencing are requested to set up in sensitive areas, such as airports, nuclear facilities, and data centers, to protect them from hostile UAV invasion.

However, the enforcement of regulations is not an easy task in practice.
Plenty of UAVs are still unregistered, and many UAVs do not have geo-fencing or the geo-fencing can be turned off easily.
There is an urgent need to quickly detect an intruder UAV in a restricted area, or assist the forensics investigation to identify its appearance and operation mode.
An ideal detection technique should give us the alert when the restricted area is invaded by unwanted UAVs at the earliest stage.
After that, counter-measures for intruder UAVs can be applied and the UAV owner may be tracked or located.
Therefore, how to efficiently detect the consumer UAVs is of utmost importance.

Other than detecting UAVs, identifying UAVs' operation mode will be very useful for forensics purposes. 
Being able to identify the operation mode of intruder UAVs can help investigators to restore the course of the incidents, which could be used as court evidences in a legal process and help law enforcement officials to improve countermeasures or responses to various possible UAV incidents.

Many physical detection mechanisms, such as radar \cite{moses2011radar, DH2017}, acoustic \cite{zelnio2009low, marmaroli2012uav}, and vision \cite{rozantsev2017detecting, PA2018, AR2015}, have been proposed for UAV detection. When using only one of these sensors for detection, these methods may get less effective in some practical scenarios, especially in a crowded urban environment.
The radar signals may get blocked by walls, buildings, and other obstacles, which are very common in a civilian environment.
The vision detection technique cannot detect the UAV in non-line-of-sight scenarios and dark.
The acoustic detection can be interfered by the environment noises which may overwhelm the relative small sound produced by tiny rotor-craft or gliding fixed-wing UAV.

Aiming to complement the above conventional physical detection mechanisms, we propose to explore machine learning-based Wi-Fi traffic identification approaches to achieve fast UAV detection and operation mode identification.
It is motivated by the observation that many existing consumer UAVs are equipped with Wi-Fi interfaces and communicate with a user handheld device (e.g., smartphone) for command control or video streaming.  
Detecting UAVs through wireless traffic identification brings us several advantages over existing mechanisms.
First, Wi-Fi signal sensing and packet capturing are less affected by obstacles, other flying objects, acoustic noise, or light conditions that could affect physical detection mechanisms.
Second, Wi-Fi data traffic provides cyber information about UAVs' type and their operation mode, which can be very useful for forensics investigation.

\emph{Challenges:} At the same time, UAV detection through Wi-Fi traffic identification introduces unique challenges that separate it from traditional traffic identification \cite{NJ2014, AMMH2004, Bar2010} and sensing tasks as follows:

1) UAV traffic can be encrypted.
Therefore, existing network monitoring and intrusion detection mechanisms that are based on packet header examination or port filtering are not applicable to encrypted UAV traffic.
For example, Wi-Fi controlled UAVs (such as DJI and Bebop drones) use WPA2 to secure the wireless communication.
Although SSID in the MAC frame may reveal information about the type or vendor of the drone, it can be easily changed through drone control apps.
2) Existing machine learning methods cannot be directly used to identify UAV traffic in a timely manner.
For real-time applications, we need to identify the UAV as soon as it is appearing in or approaching to a restricted area.
From learning and classification perspective, traditional machine learning methods \cite{NJ2014, AMMH2004} that only aim at minimizing detection error cannot be directly applied.
Detection delay introduced by the computations on feature generation and future packet arrival time should also be considered.
3) Traditional time series early detection strategies \cite{AD2015} cannot be applied to UAV traffic.
The inter-packet arrival time of UAV traffic is random, so the traditional time series early detection method which is based on fixed time intervals cannot be directly applied.

To address the above challenges, we propose a delay-aware machine learning-based UAV detection framework to strike a tunable balance between UAV detection accuracy and delay.
Our classification framework treats the encrypted data flow as a time series and extracts statistical features only based on the packet size and inter-arrival time. 
By considering the computation time among different features, our framework adopts a re-weighted $\ell_1$-norm regularization and integrates feature selection and performance optimization in one objective function.
To tackle the packet inter-arrival time uncertainty when estimating the delay cost function, we use 
maximum likelihood estimation
(MLE) method to estimate the packet inter-arrival time.
Finally, expected total cost function integrates misclassification/misdetection and delay cost which are updated online when a new packet arrives and an optimal detection decision is made to minimize the expected total cost function. 

Our main contributions are summarized as follows:
\begin{itemize}
\item We propose a machine learning-based framework to achieve delay-aware UAV detection and operation mode identification over encrypted Wi-Fi traffic.
This framework extracts features derived only from information of packet size and inter-arrival time.
This framework can be applied to other types of encrypted traffic, such as cellular traffic or proprietary protocol traffic as long as the packet size and interval can be measured.

\item In order to reduce the model prediction time for fast UAV detection, our framework adopts $\ell_1$-norm regularization and integrates feature selection and accuracy optimization in one objective function, which considers the feature importance and difference of computation time among different features. 

\item We propose to use model-based MLE method to estimate the packet inter-arrival time.
Then using the mean square error (MSE) as a well-known metric, we evaluate the performance of the estimation on the collected real-world dataset.  

\item Other than detecting and identifying different types of UAVs, our proposed method further identifies the UAV's operation mode, such as standby, hovering, flying, etc. 

\item We collect a large amount of real-world encrypted Wi-Fi data traffic of non-UAV and eight types of consumer UAVs, and conduct extensive evaluations on the performance of the proposed methods. 
\end{itemize}

Through comprehensive study, we obtain the following findings:

\begin{itemize}
\item The UAV traffic presents different patterns from non-UAV traffic. Therefore, machine learning based methods work well to differentiate UAV traffic from a wide range of non-UAV traffic. 

\item Due to vendor specific implementation of UAV command control and video streaming protocols, different types of UAVs present different traffic patterns which can be used to classify UAVs from different vendors.

\item The UAV Wi-Fi traffic presents different patterns under different UAV operation modes.
This finding implies a strong correlation or coupling between cyber information (data traffic) and physical information (operation mode) of UAVs. 
This finding is expected to motivate new cyber-physical defense and forensics mechanisms that leverage this  cyber-physical coupling.
We believe this methodology can be applied to other cyber-physical systems (CPS) and motivate more in-depth study on cyber-physical attack co-detection or co-defense for many Internet of Things (IoT) applications, such as connected cars, smart home, smart healthcare, and industrial control systems. 
\end{itemize}

\vspace{4mm}

The rest of the paper is organized as follows. 
Section \ref{RelatedWork} discusses the related works.
Problem setup is described in Section \ref{ProblemSetupNew}.
Section \ref{EarlyUAVDetection} presents our proposed method on delay-aware UAV early detection and operation mode identification.
Collection and analysis of real-world dataset is described in Section \ref{DataCollectionandAnalysis}.
Extensive performance evaluation is conducted in Section  \ref{UAVDetectionImplementation}.
Section \ref{Discussion} provides more detailed discussion on practical aspects of the proposed methodology and possible future work in this area.
Section \ref{conclusion} concludes the paper.

\section{Related Work}
\label{RelatedWork}

\subsection{UAV Detection Mechanisms}
In this section, we briefly describe the existing UAV detection methods.

\subsubsection{UAV Detection Through Physical Sensing}

Existing UAV detection mechanisms mainly focus on physical sensing through various means, including radar, vision, and acoustic.

Radar system is one of the well-known and oldest techniques in aircraft detection dating back to World War II.
In order to adapt the detection of small size UAVs, X-band radar systems were proposed \cite{MA2014, MAMR2011}.
However, in the metropolitan areas (e.g., a city) radar based detection may become less effective due to its line-of-sight requirement \cite{GMTR2016}.
The vision-based UAV detection based on video cameras \cite{GF2015} has the same weakness as radar based techniques, as it also requires line-of-sight between the camera and UAV.
However, if the cost is not a much concern, building a detection system based on multiple radars and cameras fusion to cover the targeted area would be a reliable and promising UAV detection system.

The acoustic signal-based UAV detection is a method that can solve the out-of-sight problem \cite{zelnio2009low, marmaroli2012uav, sutin2013acoustic}.
However, this method has its own drawbacks as well. 
First, the acoustic signal coming from the UAV can be quite noisy due to the noise generated at the motors of electric-powered rotor-craft with fixed wings \cite{PMXF2012}. 
Second, other similar acoustic signal generating devices such as electric weed whackers can generate sound signals quite similar to UAV's.
In order to overcome the drawbacks of individual techniques, hybrid solutions have been proposed by combining the acoustic sensor and video camera \cite{JBFP2015}.
Another hybrid solution incorporates the radar sensor as well \cite{WSGA2011}.

\subsubsection{RF-fingerprinting Based UAV Detection}
Recently, Zhao \emph{et al.} \cite{Zhao2018} proposed a new method of RF signal fingerprinting in order to detect and identify the type of the UAVs. To do that, they propose to use Auxiliary Classifier Wasserstein Generative Adversarial Networks (AC-WGANs) based on the wireless signals collected from various types of UAVs.
According to their results, this method can detect the UAVs in indoor and outdoor environments with average accuracy of 95\% and 80\%, respectively. 
In the other recent work \cite{Zhao2018}, 
Bisio \emph{et al.} \cite{Bis2018} proposed a Wi-Fi statistical fingerprint-based amateur UAV detection method by applying existing multiclass classification machine learning algorithms.
In this work the detection delay is not a concern, and thus, the main goal is to train a machine learning model to detect the intruding UAV based on the predefined and fixed number of statistical features which are computed in every fixed window size.
Our proposed method considers detection delay and strike a tunable balance between detection accuracy and delay as well as feature computation time. 
Ezuma \emph{et al.} \cite{Martin2019} proposed a new detection and classification of micro-UAVs using RF fingerprints of the signals transmitted from the controller to the micro-UAV.
In their technique, they utilized wavelet domain analysis to remove the bias in the signals which also helped in the processing data size reduction.
For the classification purposes, a naive Bayes approach has been applied to distinguish the UAV signal frames from the non-UAV classes. 
In the testing phase, a signal energy level detection is also integrated to improve the detection performance.
In average, the micro-UAV detection accuracy of 96.3\% is achieved under various signal-to-noise
ratio (SNR) levels on the channel. 
We believe integrating our work with the RF fingerprinting method proposed by Ezuma \emph{et al.} \cite{Martin2019} would result in a delay-aware, more robust, and accurate UAV detection system as each method could be a very suitable complement to the other.

A recent work \cite{SBRB2017} has been proposed to detect the approaching of a UAV within a short distance through the observation of received signal strength (RSS) changes of Wi-Fi signals.
However, an intruder UAV may just launch inside a restricted area and hover, or standby on a neighboring roof to spy on someone. 
In these scenarios, this method will not work.
Moreover, a changing RSS is not necessarily introduced by UAVs, but could be other moving objects with Wi-Fi interfaces, such as mobile users carrying smartphones or a driving car equipped with Wi-Fi connections.
So the application scenario of this proposed technique is limited. 

In another work \cite{Nguyen2017}, the authors propose a new RF-based drone detection method based on the physical characteristics of the drone, such as body vibration and body shifting, which impact the wireless signal transmitted by the drone during the communication.
This method is not useful when the UAV is in the standby mode.
Moreover, both \cite{SBRB2017} and \cite{Nguyen2017} require line-of-sight connection between the RF signal monitoring system and UAV. 
Our proposed method based on Wi-Fi traffic identification relaxes this strong assumption.

Recently, Sciancalepore \emph{et al.} \cite{Sci2019} proposed to detect drones' status, flying or laying on the ground, using encrypted network traffic identification. 
Three standard binary classification algorithms, Trees-J48 (J48), random forest (RF), and neural networks (NN) are applied to a 3DR SOLO UAV traffic dataset to identify the drone status.
Different from this work, in our work we identify eight different operation modes as well as we propose a machine learning framework which addresses the delay-aware UAV detection problem. 
Our detection system is also tested in the presence of non-UAV data traffic to detect the UAVs used in the experiment.

\subsection{Data Traffic Classification/Identification}
Classical approaches such as \emph{port-based}, \emph{payload-based} and \emph{deep packet inspection} can be used to identify the type of the non-encrypted network data traffic.
However, nowadays many application data traffic are encrypted for security purposes, and our work is closely related to encrypted data traffic classification/identification.
There are several works for identifying the encrypted data flow based on protocol data fingerprinting in wired and wireless networks, where commonly a combination of statistical and machine leaning approaches have been used \cite{NJ2014, GM2012, AMMH2004,Bar2010, TNGA2008}.

In \cite{NJ2014}, a new support vector machine (SVM) based method is proposed to identify three types of traffic, HTTP, File Transfer Protocol (FTP), and Email. 
One of the pioneering works in this area applies classification techniques to classify traffic in a wired network into classes of bulk transfer, small transactions, and multiple transactions \cite{AMMH2004}.
Bernaille \emph{et al.} \cite{Bar2010} show that it is possible to distinguish the behavior of an application from the observation of the size and the direction of the first few packets of the  Transmission Control Protocol (TCP) connection.
In this work, three classical clustering algorithms, K-Means, Gaussian Mixture Model and spectral
clustering are applied on the dataset to identify the flow.
However, this method requires packet header traces analysis, and initial TCP connection packets.
Xie \emph{et al.} \cite{GM2012} proposes a new method called subspace clustering technique (SubFlow), which learns the intrinsic statistical features of each application to classify and identify the flow. 

However, our work is different from the existing traffic identification works in the following aspects: 
1) Our model provides packet-by-packet analysis, hence the decision is made in a timely manner as packets enter the detection system. 
2) Our model adaptively finds the optimal number of the packets that are needed for optimal identification with high accuracy, while considering time cost (or delay). 
3) When training the model, feature generation time is also considered and critical features are selected. 
Therefore, in the prediction/detection, useless features are not generated, and thus the detection delay is reduced.

Compared to our earlier work \cite{CPS-Sec2019}, in this paper, operation mode of the detected UAV is identified. 
To do so, we collect a large amount of real-world operation mode data traffic of four UAVs and apply multiclass classification machine learning algorithms to identify the modes.
We also extend our delay-aware UAV detection test from four to eight commonly used consumer UAVs and conduct extensive evaluations on the performance of the proposed methods.
Moreover, we provide performance evaluation for the packet inter-arrival time estimation using MLE. 
The results indicate that mean square error (MSE) of estimation is reduced when the information of a large number of packets are available in the detection system.

Note that our detection system is applicable to detect the intruding UAVs controlled by a user handheld device (e.g., smartphone). In other words, communication link between the UAV and controller should be established in order to monitor the traffic and detect the UAV by the proposed method. On the other hand, our method will be ineffective if the intruding UAV is equipped with the advanced autonomous systems such as Autonomous Guidance, Navigation and Control (GN\&C) where no ground control station is required for command and control.

\section{Problem Setup}
\label{ProblemSetupNew}
\subsection{System Setup}
\label{ProblemSetup}
\begin{figure}[t]
			\centering
					\includegraphics[scale=0.4]{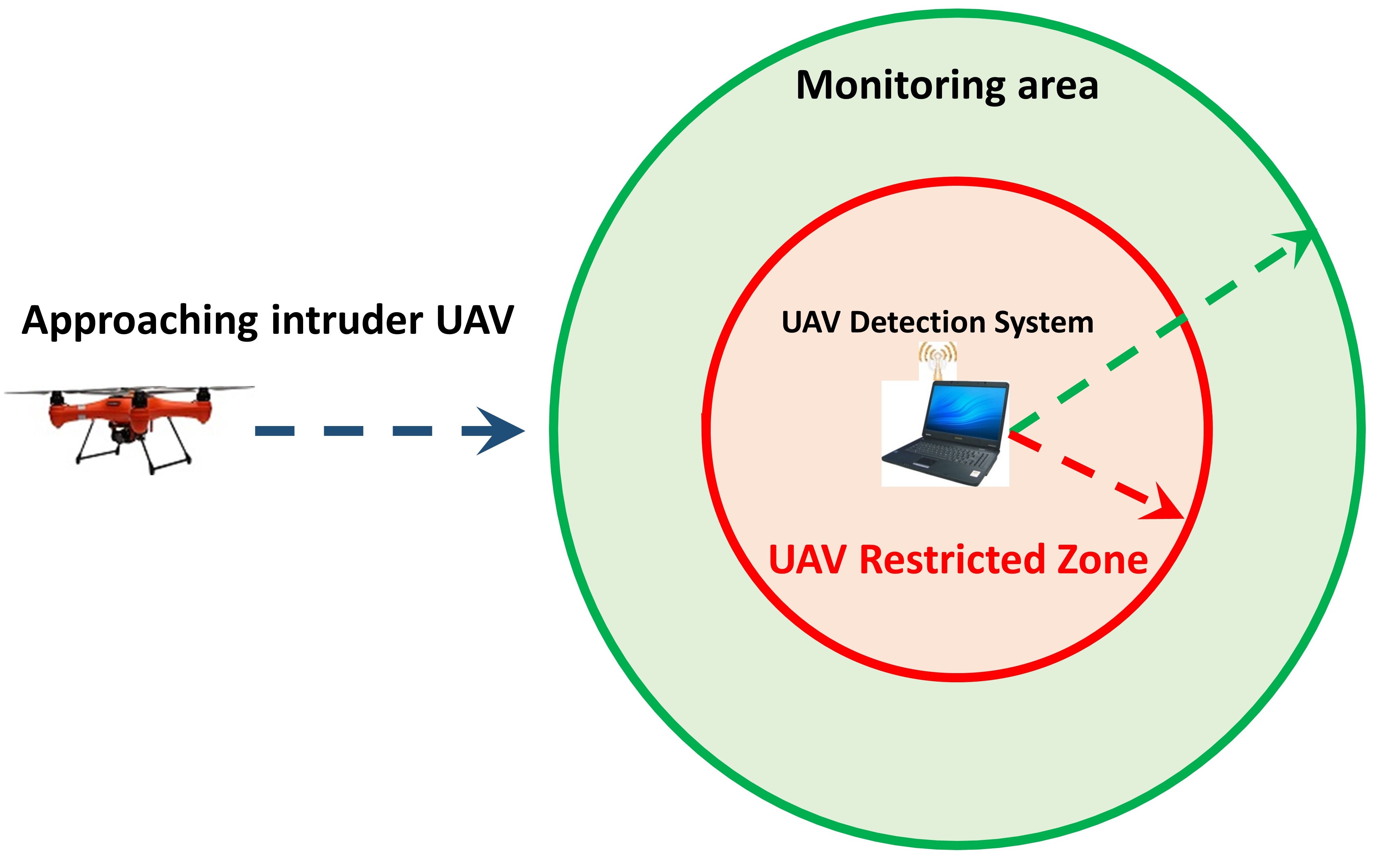}
			\caption{System model.}
							\label{systemmodel}
\end{figure}
There is a Wi-Fi signal sensing and packet capturing system that can collect all the Wi-Fi traffic within a physical sensing range in real-time.
There can be multiple Wi-Fi users in the sensing range and multiple UAVs or non-UAV devices.
The sensing system may capture non-encrypted packets, for which we assume the system can tell the application types of the corresponding flow by examining the packet headers or contents.
Since these non-encrypted packets can be easily identified by existing methods, we will only focus on encrypted Wi-Fi traffic in this paper.
For an encrypted Wi-Fi frame, the information we can obtain about the frame is its source and destination MAC addresses, transmitter and receiver MAC addresses, packet size, and packet arrival time together with other MAC header information, such as frame type (control, management, or data), sequence number, and duration/connection ID.

In this paper, we mainly focus on two sequential tasks. 
 1) Wi-Fi controlled UAV detection:
There are a large body of such kind of UAVs on the market, such as DJI, Bobop, DBPower drones,  and etc.
We assume the Wi-Fi communication between the drone and controller (e.g., smartphone) is encrypted using security protocols, such as WPA2.
We assume a drone restricted area (i.e., No Drone Zone) which \emph{no} UAVs are allowed to enter and operate (see Fig. \ref{systemmodel}). 
Our detection system can be implemented in the center of UAV restricted zone to monitor the area and detect any approaching UAVs as quickly as possible with a high accuracy. 
2) UAV operation mode identification:
For any detected UAV types in the first step, further identify its operation mode. 
Operation mode consists of standby, hover, forward, backward, and etc.
It is noted that, if in case, a malicious spying UAV just get turned on or launched inside the restricted zone and stay on in the standby or hovering mode to accomplish the spying mission, our detection system not only can detect the presence of the spying UAV, but it also can identify in which mode the spying UAV is operating.

For the UAV detection and operation mode identification, we only use data frames.
We divide the encrypted Wi-Fi traffic into individual flows according to the pair of source and destination MAC addresses.
A unique flow includes the packets between a pair of nodes.
The traffic in a flow can be bi-directional or unidirectional.
In a  real-time scenario, these flows usually interleave with each other in time.
The goal of this paper is to identify the UAV data flows when frames are captured and decide the UAV type and its operation mode in a quick manner with high accuracy.

\subsection{Delay-aware UAV Detection Problem Formulation}
\label{sec:preliminaries}
The UAV detection over encrypted Wi-Fi traffic can be formulated as a machine learning classification problem.
Let's assume that we can obtain a large training dataset with $m$ flow traces with each trace having $n$ consecutive packets. 
The traces contain UAV and non-UAV flows which are labeled with their corresponding flow types $y_i \in \Yb$, where $\Yb = \left\{\text{UAV}_1,\text{UAV}_2,...,\text{UAV}_{\upsilon-1},\text{non-UAV}\right\}$ 
and $\upsilon = |\Yb|$ denotes the number of class types in set $\Yb$.
$\text{UAV}_j$ for $j \in \left\{1,...,\upsilon-1\right\}$ denotes the UAV type $j$. 

\emph{Packet size} and \emph{packet inter-arrival time} are two key attributes we extract from these traces for UAV detection and operation mode identification.
The sequences of packet size and packet inter-arrival time for the $i$th trace are denoted by $\mathbf{x}_i$ and $\boldsymbol{\tau}_i$, respectively. 
Now, let $\mathbf{x}_i=(x_{i,1},\cdots,x_{i,n})$, where $x_{i,j}$ for $i= 1,...,m$ and $j = 1,...,n$ indicates the size of the $j$th packet in the $i$th trace. Similarly, let $\boldsymbol{\tau}_i=(\tau_{i,1},\cdots,\tau_{i,n})$, where $\tau_{i,j}$ denotes the inter-arrival time between the $j$th and $(j+1)$th packets ($j<n$) in the $i$th trace.
Define a finite set $S = \left\{((\mathbf{x}_i,\boldsymbol{\tau}_i),y_i)\right\}_{i \in \left\{1,...,m\right\}}$ where the pair $(\mathbf{x}_i,\boldsymbol{\tau}_i)$ represents the packet size and inter-arrival time of the $i$th trace in set $S$, respectively.

Let $\mathbf{\tilde{x}}(t_k)$ denotes the received incoming traffic up to its $k$th packet arrived at time $t_k$.
Assume a set of multiclass classifiers $\mathcal{H} = \left\{h^j_\gamma\right\}_{j \in \left\{1,...,n\right\}}$ are trained to classify the incoming traffic flow $\mathbf{\tilde{x}}(t_k)$, where 
$\gamma \in \Yb$. 
When the Wi-Fi sensing system receives a new packet of the incoming traffic flow $\mathbf{\tilde{x}}(t_k)$, its new features are extracted and incorporated in the prediction system. 
Intuitively, as more packets arrive, more accurate information about the traffic can be gained.
On the other hand, collecting more packets introduces longer identification delay. 
Therefore, there is a trade-off between detection accuracy and delay in the detection process.

Let $C_1(\hat{y},\tilde{y}): \Yb \times \Yb \rightarrow \mathbb{R}$ denotes the test misclassification cost function where ($\hat{y}$, $\tilde{y}) \in \Yb$, and $\hat{y} = h_\gamma^j(\bold{\tilde{x}}(t_j))$ is the predicted class label, while the true class label of the incoming flow is $\tilde{y}$. 
Let $C_2(t_p) \in \mathbb{R}$, $p>k$, be the time cost function which indicates the time cost value if UAV detection 
is postponed up to time instant $t_p$. 
Thus, the estimated total cost function is given by
\begin{figure}[t]
			\centering
					\includegraphics[scale=1.1]{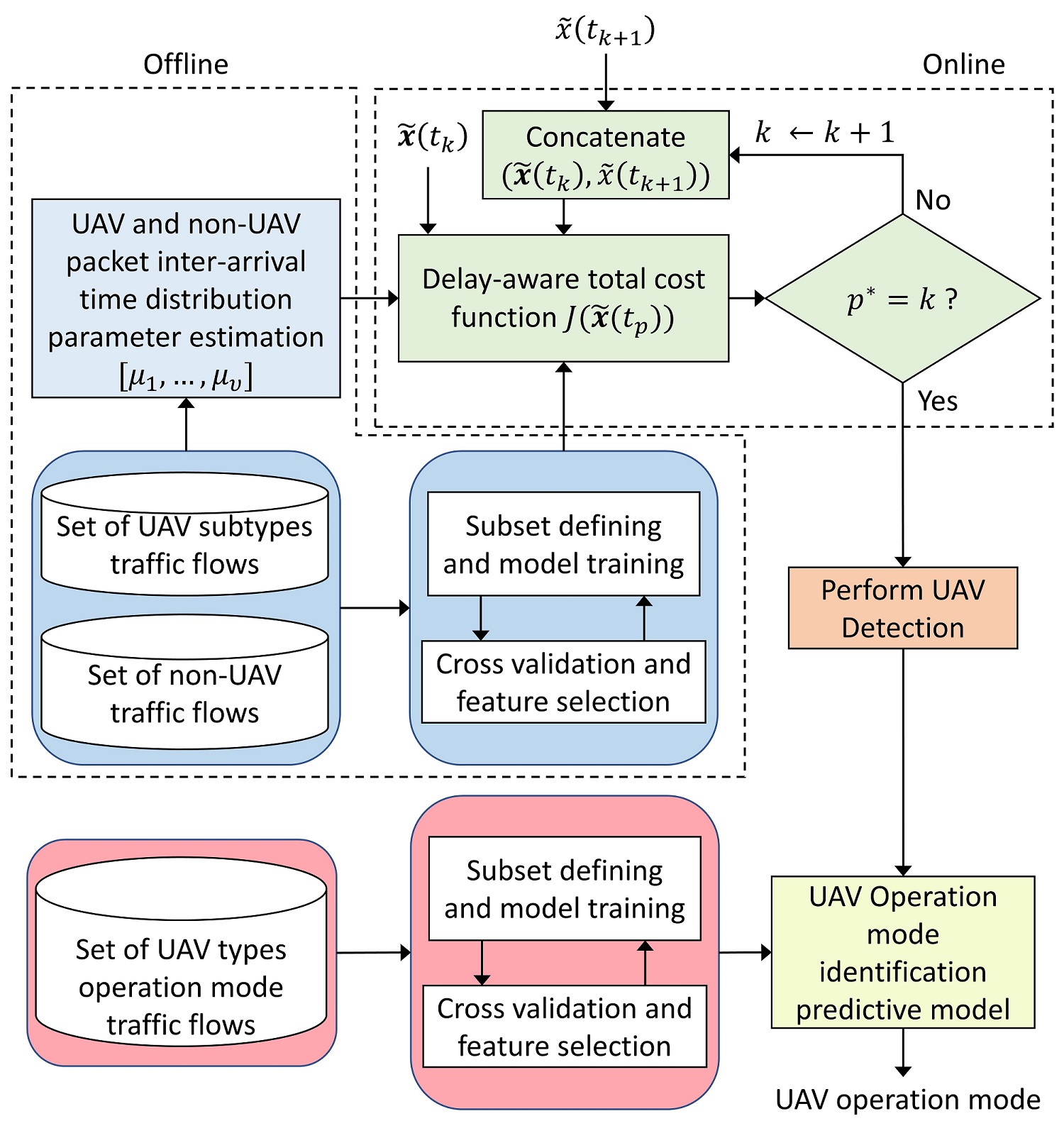}
			\caption{Delay-aware UAV detection and operation mode identification workflow.}
				\squeezeup
							\label{costfig}
\end{figure}

\begin{equation}
\begin{aligned}
\label{equ1}
J\big(\bold{\tilde{x}}(t_k)\big) = C_1\big(h_\gamma^k (\bold{\tilde{x}}(t_k),\tilde{y})\big) + C_2(t_k).
\end{aligned}
\end{equation}

In order to find an optimal trade-off between the detection accuracy and delay, the total cost function $J$ needs to be minimized. 
Hence, we formulate the delay-aware UAV detection optimization problem as follows:
\begin{equation}
\begin{aligned}
\label{equ2}
p^* = \underset{p \in \left\{k,...,n\right\}}{\arg\min} \quad J\big(\bold{\tilde{x}}(t_p)\big),
\end{aligned}
\end{equation}
where $p^*$ indicates the optimal number of the packets that needs to be received from the incoming flow before performing the UAV detection decision, and $t_{p^*}$ denotes the $p^*$th  packet arrival time. 
When $p^* = k$ is the solution of the optimization problem in (\ref{equ2}), there is no need to collect more packets because $J\big(\bold{\tilde{x}}(t_k)\big) < J\big(\bold{\tilde{x}}(t_q)\big)$ for $q = k+1,...,n$. 
Therefore, the detection is performed instantly at time $t_k=t_{p^*}$.
However, when $p^* \neq k$ is the solution of (\ref{equ2}), it means that the total cost is minimized for $k+1 \leq p^* \leq n$, thus the detection process is deferred to collect more packets from incoming flow.
For the notation purposes, let the indicator function $\mathbb{I} (p^*) = 1$ when $p^* = k$; and otherwise $\mathbb{I} (p^*) = 0$, where $p^* = k$ corresponds to the UAV detection at time $t_{p^*}$.

\subsection{UAV Operation Mode Identification Problem}
In the second task, for the detected UAV, we 
further identify the UAV's operation mode. 
Eight UAV operation modes are labeled as $ \Zb = $\{``Standby'', ``Hover'', ``Forward'', ``Backward'', ``Up'',  ``Down'', ``Right'', ``Left''\}.
In this problem, using existing machine learning algorithms a set of multiclass classifiers are trained  packet-by-packet on a real-world dataset to minimize the total operation mode misidentification cost.

\section{Delay-aware UAV Detection and Operation Mode Identification}
\label{EarlyUAVDetection}

In this section, we propose a delay-aware learning-based predictive model in order to solve the problem formulated in (\ref{equ1}) and (\ref{equ2}).
Then, we extend our work to further identify the operation mode of the detected UAV.
Fig.~\ref{costfig} illustrates the main workflow of the proposed method.

\subsection{Learning-Based Model Design}
\label{Learning-BasedModelDesign}

Based on the definition of traffic flow dataset $S$ in Section \ref{sec:preliminaries}, we further define the data of \emph{incomplete traffic flow}, where only the first  $j \leq n$ packets are available in the dataset, denoted as $S^j=\{((\bold{x}_i^j,\boldsymbol{\tau}_i^j),y_i))\}_{i \in \left\{1,...,m \right\} , j \in \left\{1,...,n\right\}}$ where $\bold{x}_i^j =  (x_{i,1},x_{i,2}...,x_{i,j})$ and $\boldsymbol{\tau}_i^j = (\tau_{i,1},\tau_{i,2},...,\tau_{i,j})$ indicate the sequence of packet size and inter-arrival time of the $i$th trace in the $j$th subset, respectively.
Next, for each dataset $S^j$, we generate a design matrix of $\bold{X}^j = [X_1^j, X_2^j ,. . ., X_m^j]^T \in \mathbb{R}^{m \times 2l}$, where $X_i^j \in \mathbb{R}^{2l}$ is a row vector as
$X_i^j=\left[V_1(\bold{x}_i^j),...,V_l(\bold{x}_i^j),V_1(\boldsymbol{\tau}_i^j),...,V_l(\boldsymbol{\tau}_i^j)\right]$
where $V_1(\cdot),...,V_l(\cdot)$ are functions which compute the statistical features of the input samples (i.e., $\bold{x}_i^j$, $\boldsymbol{\tau}_i^j$ ).
$l$ denotes the number of features.  
A list of statistical feature functions with their associated computation formula is shown in Table~\ref{tab:feature}.

\begingroup
\setlength{\tabcolsep}{0.8pt} 
\renewcommand{\arraystretch}{1.2} 
\begin{table}[t]
\scriptsize
\begin{threeparttable}
\centering
\caption{\small Statistical features (sample size $N=100$).
}
\begin{tabular}{l|c|c}
\hline
Function: Feature Name	&	Description	 & Comput. time \\
\hline
\tabincell{c}{$V_1(x)$: mean}	&	\(\bar{x}=\frac{1}{N}\sum_{i=1}^N x(i)\) & 0.672 $\mu s$
	\\
\hline
 \tabincell{c}{$V_2(x)$: median}	&		 The  higher half value of a data sample. & 4.365 $\mu s$ \\
\hline
 \tabincell{c}{$V_3(x)$:  MedAD\footnotemark[1]}&	\( MedAD=median(|x(i)-median(x)|)\) & 8.346 $\mu s$ \\
\hline
 \tabincell{c}{$V_4(x)$: STD\footnotemark[2]}	& 	\(\sigma=\sqrt{\frac{1}{N-1}\sum_{i=1}^N (x(i)-mean(x))^2}\) & 1.608 $\mu s$ \\
\hline
\tabincell{c}{$V_5(x)$: Skewness}	&	 \(\gamma=\frac{1}{N}\sum_{i=1}^N ({x(i)-mean(x)})/{\sigma})^3\)  & 14.917 $\mu s$ \\
\hline
 \tabincell{c}{$V_6(x)$: Kurtosis}	&
 \(\beta=\frac{1}{N}\sum_{i=1}^N ({x(i)-mean(x)}/{\sigma})^4\) & 14.095 $\mu s$ \\
\hline
\tabincell{c}{$V_7(x)$: MAX}	& \(H=(Max(x(i))|_{i=1...N})\) & 0.464 $\mu s$ \\
\hline
\tabincell{c}{$V_8(x)$: MIN}&	
 \(L=(Min(x(i))|_{i=1...N})\)	& 0.652 $\mu s$   \\
\hline
\tabincell{c}{$V_9(x)$: Mean Square}&
\(MS=\frac{1}{N}\sum_{i=1}^N (x(i))^2\)	& 1.147 $\mu s$ \\
\hline
 \tabincell{c}{$V_{10}(x)$: RMS} & 
 \(RMS=\sqrt{ms(x)}\) 	& 1.273 $\mu s$	\\
\hline
 \tabincell{c}{$V_{11}(x)$: PS\footnotemark[3]}	&
  \(3(mean(x)-median(x))/\sigma\) & 8.011 $\mu s$ \\
\hline
\tabincell{c}{$V_{12}(x)$: MAD\footnotemark[4]}	&
 \(MAD = \frac{1}{N} \sum_{i=1}^N |(x(i)-mean(x))|\) & 2.531 $\mu s$ \\
\hline
\end{tabular} 
\label{tab:feature}
\begin{tablenotes}\scriptsize  
\item[1] MedAD: median absolute deviation
\item[2] STD: standard deviation
\item[3] PS: Pearson skewness
\item[4] MAD: mean absolute deviation
\squeezeup
\end{tablenotes}
\end{threeparttable}
\end{table}
\endgroup

\textbf{Feature selection based on re-weighted $\ell_1$-norm by considering both feature discriminative power and computation time cost:}
Different features have different significance; we use $W^j=\{W_1^j,W_2^j,\cdots,W_{2l}^j\} \in \mathbb{R}^{2l}$ to denote the weight vector of $j$th subset for all the $2l$ features. 
Therefore, $W_i^j=0$ means that the $i$th feature of $j$th subset is not useful and can be discarded. 
Then, given the $j$th design matrix $\bold{X}^j$ and $y \in \Yb$, 
our problem is to learn a predictive mapping $h_\gamma^j(X_i^j , W^j)\rightarrow y_i$ for $i=1,2,\cdots,m$ such that: 
1) the disagreement between $h_\gamma^j(X_i^j, W^j)$ and $y_i$ is minimized; and 
2) the number of non-zeros in $W^j$ is minimized.

Other than the fact that different features have different discriminative power for identification, we also observe that different features consume different amounts of computation time (refer to Table~\ref{tab:feature}).
Therefore, in the $j$th subset among the features that bring the same discriminative power in UAV identification, we tend to remove the one(s) that consume more time.
This requires us to give personalized penalty on each different feature. 
The more time one consumes, the more penalty it is given. 
Thus, instead of using conventional $\ell_1$-norm regularization \cite{FBRJ2012} 
that penalizes all the features evenly, we propose the following new objective loss function with the re-weighted $\ell_1$-norm:
\begin{equation}
\label{eq:objfun2}
\min_{W^j} \; \mathcal{L}\big(y,h_\gamma^j(\bold{X}^j ,W^j)\big)+\sum_{i=1}^{2l}\lambda_i^j|W_i^j|,
\end{equation}
where the strength of penalty $\lambda_i^j$ for $i$th feature in $j$th subset is proportional to the computational time for this feature.
Therefore, the objective function in (\ref{eq:objfun2}) will minimize  the misclassification error and enforce some $W_i^j$'s to be zeros, especially those that consume  more computation time. 
Since for computing of expected misclassification cost function, we shall need probabilistic output of the classifier, then we choose $h_\gamma^j(\cdot)$ to be one-vs-all logistic regression function 
\cite{murphy2013machine}.
One-versus-all logistic regression is a generalized version of the logistic regression into multiclass classification.
\floatname{algorithm}{Algorithm}
\renewcommand{\algorithmicrequire}{\textbf{Input:}}
\renewcommand{\algorithmicensure}{\textbf{Output:}}

\begin{algorithm}[t]
    \caption{\small Training phase framework for UAV identification}
    \label{alg:optimizatiomn}
 \begin{algorithmic}
 {\scriptsize
   \REQUIRE Wi-Fi traffic trace dataset $\left\{(\bold{x}_i(t_n),y_i)\right\}$, $(y_i,\gamma) \in \Yb$,\\
\hspace{6mm}   $\upsilon = |\Yb|$, $i \in \left\{ 1,...,m \right\}$, $j \in \{1,...,n\}$;
   \ENSURE $E^j(X_i^j)$, $W^j$, set of classifiers $\mathcal{H} = \left\{h_\gamma ^ j\right\}$;
\begin{steps}
\setlength{\itemindent}{1mm}
\setlength\itemsep{0.5em}
\item Extract packet size and inter-arrival time of encrypted Wi-Fi traffic traces and create dataset $S = \left\{((\bold{x}_i,\boldsymbol{\tau}_i),y_i))\right\}$;
\item Define subsets $S^j=\left\{((\bold{x}_i^j,\boldsymbol{\tau}_i^j),y_i))\right\}$, where\\  
\hspace*{0mm} $\bold{x}_i^j =  (x_{i,1},x_{i,2}...,x_{i,j})$ and $\boldsymbol{\tau}_i^j = (\tau_{i,1},\tau_{i,2},...,\tau_{i,j})$;
\item Determine design matrices $\bold{X}^j = [X_1^j, X_2^j ,. . ., X_m^j]^T$, where \\
$X_i^j=\left[V_1(\bold{x}_i^j),...,V_l(\bold{x}_i^j),V_1(\boldsymbol{\tau}_i^j),...,V_l(\boldsymbol{\tau}_i^j)\right] \in \mathbb{R}^{2l}$;
\item  Train a set of classifiers $\mathcal{H} = \left\{h_\gamma^j\right\}$ by solving (\ref{eq:objfun2}):\\
\hspace*{16mm} $\min\limits_{W^j} \; \mathcal{L}\big(y,h_\gamma^j(\bold{X}^j , W^j)\big)+\sum_{i=1}^{2l}\lambda_i^j|W_i^j|$;\\
\item  Compute expected training misclassification function: \\
\hspace*{0mm}       \textbf{for} $i \in \left\{1:m\right\}$\\
\hspace*{4mm}       \textbf{for} $j \in \left\{1:n\right\}$\\
\hspace*{8mm}			\textbf{for} $\gamma \in \left\{\text{UAV}_1,\text{UAV}_2,...,\text{UAV}_{\upsilon-1},\text{non-UAV}\right\}$\\
\hspace*{13mm} Compute $P_{j}(\hat{y}=\gamma|X_i^j;W^j) = h_\gamma ^ j (X_i^j , W^j);$ \\
\hspace*{13mm}				Compute $E^j(X_i^j)$ using (\ref{equ77}).
\end{steps}}
\end{algorithmic}
\end{algorithm}

Next, we compute the training expected misclassification cost function on each trace $i$ for every subset $j$ as follows:
\begin{equation}
\begin{aligned}
\label{equ77}
E^j(X_i^j) = \sum_{y_i \in \Yb} P(y_i|X_i^j) &\sum_{\hat{y} \in \Yb} P_{j}(\hat{y}|X_i^j;W^j) C^j(\hat{y}|y_i),\\
\end{aligned}
\end{equation}
where $P(y_i|X_i^j) = 1$ if $\hat{y}=y_i$ and $0$ otherwise. 
Then, a set of one-vs-all logistic regression classifiers $\mathcal{H} = \left\{h_\gamma^j\right\}$ for $j= 1,...,n$ and $\gamma \in \left\{\text{UAV}_1,\text{UAV}_2,...,\text{UAV}_{1-\upsilon},\text{non-UAV}\right\}$ are trained.
Based on the probabilistic output of one-vs-all logistic function, we can compute  $P_{j}(\hat{y}=\gamma|X_i^j;W^j)=h_\gamma^j(X_i^j,W^j)$.
$C^j(\hat{y}|y_i)$ denotes the misclassification cost function of training dataset. 
$C^j(\hat{y}|y_i)$ = 1 if $\hat{y}=y_i$ and $0$ otherwise, and $\hat{y} = \max\limits_{\gamma} h_\gamma^j(X_i^j,W^j)$.
We summarize the model training phase for UAV identification in Algorithm 1. 

\subsection{Delay-aware Predictive Model}

\subsubsection{{Expected missclassification cost function $C_1$}}
In the prediction phase of the incoming flow $\bold{\tilde{x}}({t_k})$, in order to compute the expected misclassification cost function $C_1$, $E^j(X_i^j)$ is weighted based on the incoming traffic's Euclidean distance 
from every trace in the training dataset.
Consider $\bold{\tilde{x}}(t_k)$ be the incoming flow and $\tilde{X}^k \in \mathbb{R}^{2l}$ its corresponding feature values.  
The weight function is defined as a normalized sigmoid function by $f_{w_i}^k = s_i^k/\sum_{i}^{m} s_i^k$ where $s_i^k = 1/1+exp^{-\eta \Delta_i^k}$, and
$\eta$ is some positive constant, and $\Delta_i^k = {\bar{D}_{i} - d_i^k}/\bar{D}_{i}$ is the normalized average distances between $\tilde{X}^k$ and all the traces in the training dataset \cite{AD2015}.
$d_i^k = ||\tilde{X}^k - X_i^k||_2$ indicates the
Euclidean distance of the incoming flow from $i$th trace in the dataset.
In fact, the weight function $f_{w_i}^k $ plays the role of a \emph{similarity function} which measures how close the incoming traffic flow is to each of the traces in the training dataset.
Hence, the expected misclassification cost function for $\bold{\tilde{x}}(t_k)$ is defined as follows:
\begin{equation}
\begin{aligned}
\label{equ100}
C_1\big(h_\gamma^k (\bold{\tilde{x}}(t_k),\tilde{y})\big)= \sum_{i=1}^{m} f_{w_i}^k E^k(X_i^k). 
\end{aligned}
\end{equation}
The above equation indicates that more weights are multiplied to the training expected misclassification value of the $i$th trace if its distance from the incoming flow is larger and vice versa.

\subsubsection{{Estimated time cost function $C_2$}}
\begin{figure}[t] 
		  \centering
	 	\includegraphics[scale=0.8]{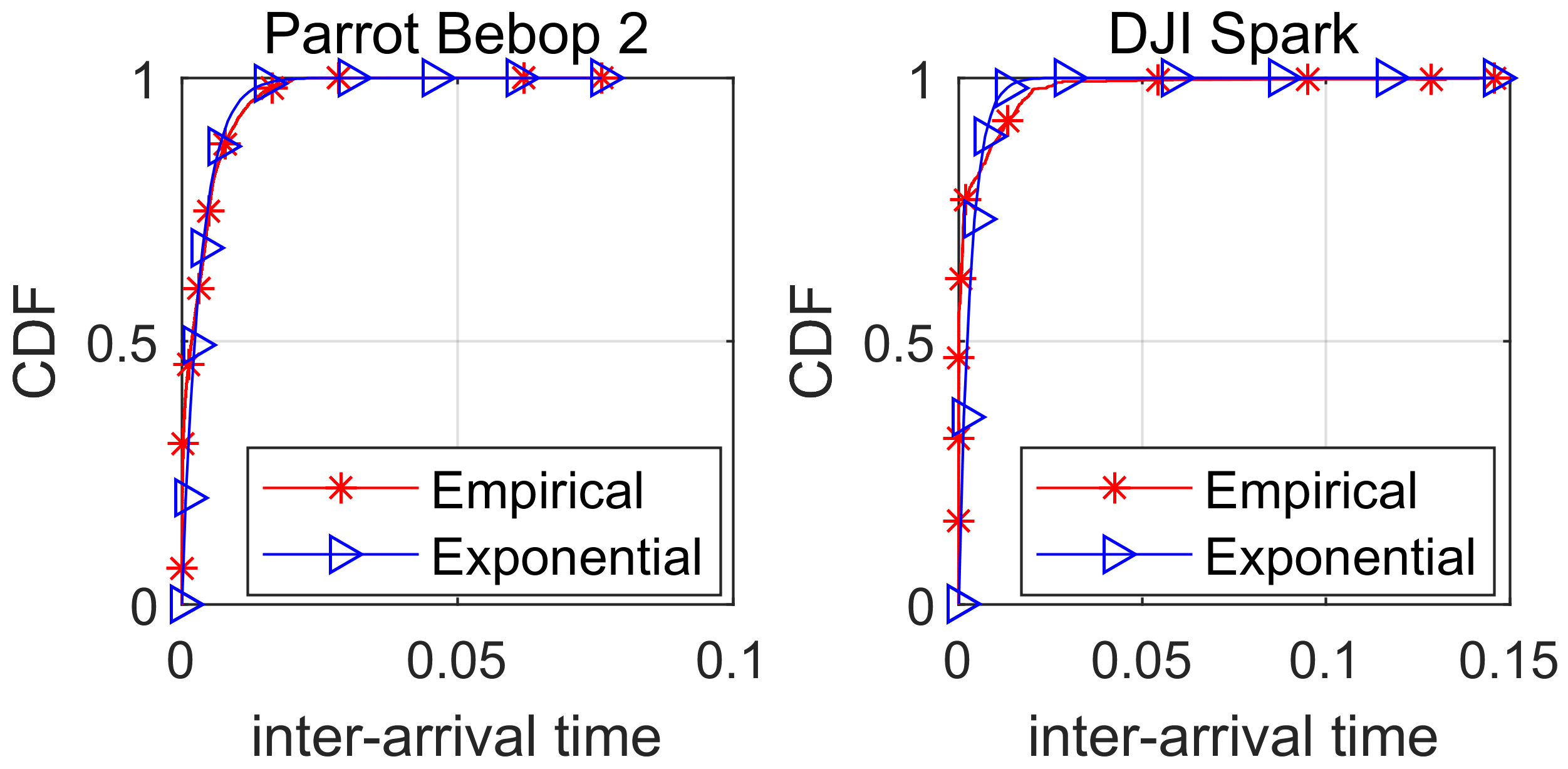} 
				\caption{CDF plot to compare the fit of exponential distribution to the empirical CDF of packet inter-arrival time}.
       \label{distfig}
 \end{figure}

For the incoming flow $\bold{\tilde{x}}(t_k)$, future packet arrival times are unknown and random.
This uncertainty in packet arrival times introduces difficulties in constructing a delay-aware UAV identification algorithm.
In order to tackle this challenge, we propose to estimate the incoming flow's future packet inter-arrival time according to the exponential distribution with parameter $\hat{\mu}_i$ for $i=1,...,\upsilon$ achieved by MLE method \cite{AK2010}.
In Table \ref{tab:good}, we present the goodness-of-fit statistics to show that exponential distribution provides a good approximation for packet inter-arrival time estimation. 
For the illustration purposes, we also graphically show the goodness-of-fit for Bebop 2 and DJI Spark in Fig.~\ref{distfig}. 

\begingroup
\begin{table}[t]
\begin{centering}
\scriptsize
\caption{\small Empirical and exponential cumulative distribution function (CDF) goodness-of-fit statistics.}

\begin{tabular}{c|c|c|c|c|}
\cline{2-5}  & \multicolumn{4}{ c| }{UAV type} \\ 
\cline{2-5}  & Bebop 1 & Bebop 2 & Spark & UDI  \\ 
\cline{1-5} \multicolumn{1}{ |c|  }{KS\footnotemark[1]} & 0.0612 & 0.0508 & 0.0773 & 0.0811\\
\cline{1-5} \multicolumn{1}{ |c|  }{CvM\footnotemark[2]} & 0.0720 & 0.0633 & 0.0691 & 0.0794 \\ 
\cline{1-5}
\cline{2-5}  & \multicolumn{4}{ c| }{UAV type} \\
\cline{2-5}  & Discovery & Tello & TDR & Wingstand  \\ 
\cline{1-5} \multicolumn{1}{ |c|  }{KS} & 0.0801 & 0.0622 & 0.0910 & 0.06741\\
\cline{1-5} \multicolumn{1}{ |c|  }{CvM}    & 0.0865 & 0.0890 & 0.0533 & 0.0695 \\ 
\cline{1-5}
\end{tabular}%
\label{tab:good}
\begin{tablenotes}\scriptsize  
\item \hspace{7mm} $^1$ KS: Kolmogorov-Smirnov
\item \hspace{7mm}  $^2$ CvM: Cramer-von Mis
\end{tablenotes}
\end{centering}
\end{table}
\endgroup

\begin{algorithm}[t]
    \caption{\small Delay-aware UAV identification}
    \label{alg:optimizatiomn}
 \begin{algorithmic}
 {\scriptsize
  \REQUIRE Incoming traffic flow $\bold{\tilde{x}}({t_k})$, $E^j(X_i^j)$, $W^j$, \hspace*{0mm} $\mathcal{H} = \left\{h_\gamma^j\right\}$,\\
\hspace*{6mm}$i \in \left\{ 1,...,m \right\}$, $j \in \{1,...,n\}$, $\gamma \in \Yb$,
\hspace*{0mm} $(\mu_1,...,\mu_{\upsilon})$;
\vspace{1.5mm}
\ENSURE $t_{p*}$, $p^*$, $\hat{y} = \max\limits_{\gamma} h_\gamma^k(\tilde{X}_i^k,W^k)$;
\begin{steps}
\setlength{\itemindent}{1mm}
\setlength\itemsep{0em}
\vspace{0mm}
\item Extract packet sizes $\bold{\tilde{x}}^k$ and inter-arrival times $\boldsymbol{\tilde{\tau}}^k$ of $\bold{\tilde{x}}(t_k)$, then compute statistical feature values $\tilde{X}^k$;
\item Compute $\Delta_i^k$ and weight function $f_{w_i}^k$;
\item Compute $C_1$ using (\ref{equ100});
\item Identify the trace label which has minimum distance with $\tilde{X}^k$.
Pick the corresponding class inter-arrival time from $[\mu_1,...,\mu_{\upsilon}]$, then compute $C_2$ using (\ref{equ200});
\item  Calculate expected total cost function $J$ using (\ref{equ1});
\item Compute (\ref{equ2}), if $p^* = k$ then, 
$\mathbb{I}(p^*) \leftarrow 1$ (perform UAV detection)
and break; otherwise $k \longleftarrow k+1$ and go to \textbf{Step 1};
\end{steps}}
\end{algorithmic}
\end{algorithm}


\textbf{Exponential distribution parameter estimation using MLE:}
Let $\{\Tb_{n}^i\}$ be a sequence of $n$ independent and identically distributed (i.i.d.) exponential random variables.
Thus, $\Tb_j^i \sim \text{Exp}(\mu_i)$ has a probability density function (pdf) of $f_{\Tb^i} (\tau_j^i) = \mu_i \text{exp}(-\mu_i \tau_j^i)$ for $\tau_j^i \geq 0$ with parameter $\mu_i$, where $j=1,...,n$, $i=1,...,\upsilon$ and $\upsilon = |\Yb|$.
Given the data sequence $\{\Tb_{n}^i\}$, our goal is to estimate the average packet inter-arrival time (i.e., $\mu_i$). 
Since $\Tb_j^i$ for $i=1,...,\upsilon$ and $j=1,...,n$ are assumed to be i.i.d., then the likelihood function is given by
\begin{align}
\label{equ31}
\mathcal{L}(\mu_i ; \tau_1^i,...,\tau_n^i) = \prod_{j=1}^n f_{\Tb^i} (\tau_j^i ;\mu_i )  = \mu_i^n\text{exp}\Bigg(-\mu_i \sum_{j=1}^n \tau_j^i \Bigg).
\end{align}
By taking logarithm of both sides in (\ref{equ31}), we obtain the log-likelihood function as 
\begin{align}
\label{equ34}
\boldmath{l}(\mu_i ; \tau_1^i,\tau_2^i,...,\tau_n^i) = n\text{ln}(\mu_i) -\mu_i \sum_{j=1}^n \tau_j^i.
\end{align}
Then, maximum log-likelihood estimation of $\mu_i$ is achieved by solving the first order maximization problem of 

\begin{align}
\label{equ999}
\hat{\mu}_i=\text{arg} \underset{\mu_i}{\text{max}}\ \boldmath{l}(\mu_i ; \tau_1^i,\tau_2^i,...,\tau_n^i),
\end{align}
as $\frac{d}{d\mu_i}\boldmath{l}(\mu_i ; \tau_1^i,\tau_2^i,...,\tau_n^i) = 0$, which results in 
\begin{align}
\label{equ35}
\hat{\mu}_i = \frac{n}{\sum_{j=1}^n \tau_j^i} \quad \text{for} \quad  i=1,...,\upsilon.
\end{align}

Next, we estimate the packet inter-arrival time of the incoming traffic flow $\bold{\tilde{x}}(t_k)$, through the following steps:
First, the Euclidean distance between $\bold{\tilde{x}}(t_k)$ and each trace in the training set is computed.
Second, the class label of the trace which has a minimum distance from the incoming flow is identified.
Third, the average inter-arrival time of the identified class is selected from (\ref{equ35}) to estimate the packet inter-arrival time of $\bold{\tilde{x}}(t_k)$ using the exponential distribution.

Now, let $\tilde{\tau}_{i+1}=t_{i+1}-t_i$ for $i=k,...,n$ be the packet inter-arrival time of the $\bold{\tilde{x}}(t_k)$ estimated by the above steps. Then, the estimated time cost function is obtained as
\begin{equation}
\begin{aligned} 
\label{equ200}
C_2(t_p) = \sum_{i=k}^p \tilde{\tau}_{i+1} \quad \text{for} \quad p=k,...,n
\end{aligned}
\end{equation}
where $C_2(t_p)$ is a strictly increasing function.

\subsubsection{{MLE performance metric}}
We use MSE metric to measure the performance of the parameters estimated by the MLE. 
Considering an incoming traffic flow $\bold{\tilde{x}}(t_k)$ and letting $\tau_{i+1}$ for $i=k,...,n$ be the true packet inter-arrival time of $\bold{\tilde{x}}(t_k)$, we have
\begin{equation}
\begin{aligned}
\label{equ121}
MSE_p = \frac{1}{n-p} \sum_{i=p}^{n} (\tau_{i+1} - \tilde{\tau}_{i+1})^2 \quad \text{for} \quad p=k,...,n
\end{aligned}
\end{equation} 
where $MSE_p$ denotes the MSE estimation of packet inter-arrival time of $\bold{\tilde{x}}(t_k)$ when $p$th packet arrives. 

\subsubsection{{Estimated expected total cost function $J$}}
According to (\ref{equ1}), the total cost function $J$ is defined based on $C_1$ and $C_2$ which can be computed using \eqref{equ100} and \eqref{equ200}, respectively. 
Algorithm 2 summarizes the total cost function estimation and incoming traffic flow's identification phase.


\subsection{UAV Operation Mode Identification}
We identify eight common operation modes for consumer UAVs in the market which are labeled as $\Zb = $\{``Standby'', ``Hover'', ``Forward'', ``Backward'', ``Up'',  ``Down'', ``Right'', ``Left''\}.
UAVs' operation mode is based on the type of the command they receive from the controller.
Each UAV operation mode produces a distinct traffic pattern in the Wi-Fi network. 
This pattern depends on the type of the command issued by the controller which governs different packet size and inter-arrival time in the trace.
Therefore, a multiclass classification model trained on a suitable dataset can identify a UAV's operation mode.
Given a dataset which contains a specific UAV's Wi-Fi traffic traces labeled with the operation modes mentioned in set $\Zb$, two well-recognized multiclass classification algorithms, SVM and random forest (RF) are applied to create the discriminative model.
Then, the incoming traffic flow $\bold{\tilde{x}}({t_k})$ is provided as an input to the corresponding multiclass classifier to identify the operation mode of the detected UAV. 

 \begin{figure}[t]
		\centering
			\includegraphics[scale=0.75]{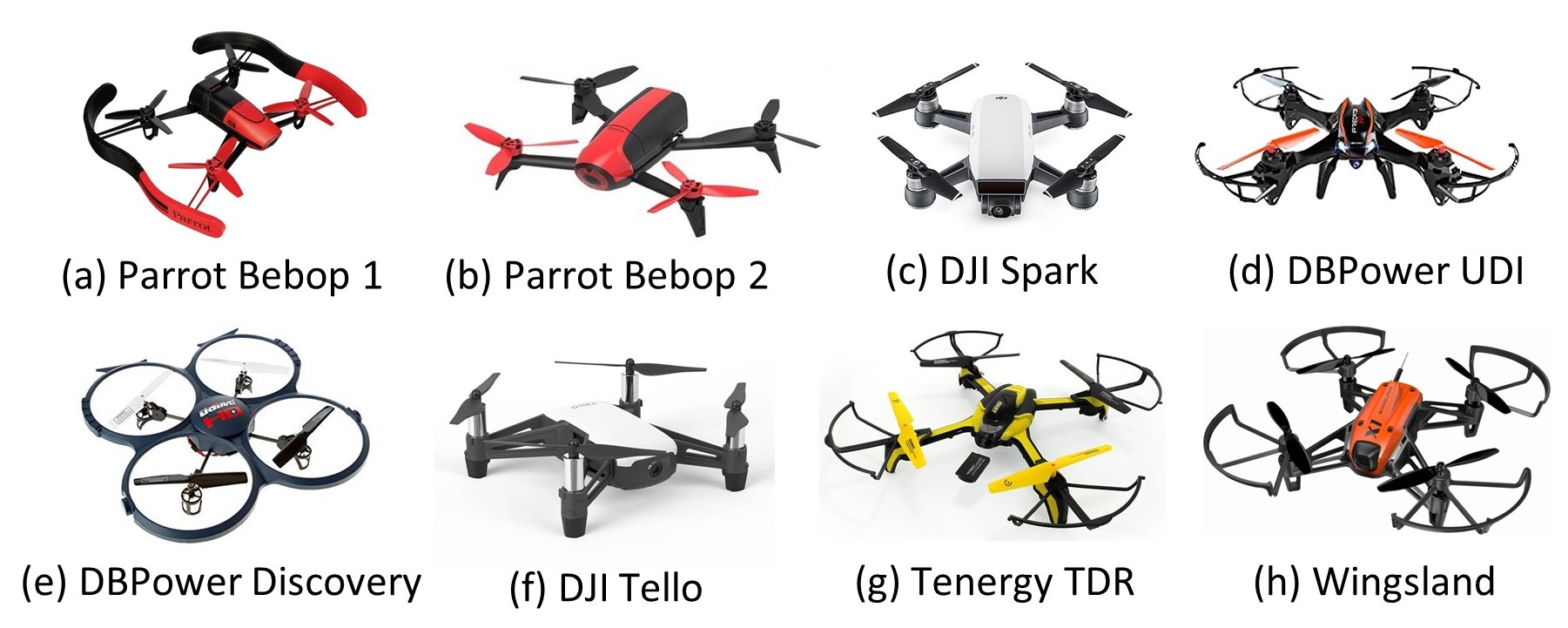}
			\caption{UAV types used in the experiments.}
			\squeezeup

					\label{UAVsubtypesfig2}
\end{figure}

\section{Data Collection and Preparation}
\label{DataCollectionandAnalysis}

\subsection{UAV Detection Dataset}

We collect traffic flows from eight types of consumer UAVs shown in Fig. \ref{UAVsubtypesfig2}:
Parrot Bebop 1 Quadcopter Drone (Bebop 1),
Parrot Bebop 2 Quadcopter Drone (Bebop 2),
DJI Spark (Spark),
DBPower UDI U842 Predator FPV (UDI),
DBPOWER Discovery FPV (Discovery),
DJI Tello (Tello),
Tenergy TDR Phoenix Mini RC Quadcopter Drone (TDR), and
Wingsland Mini Racing Drone (Wingsland). 
We use a DELL Latitude laptop embedded with a wireless network interface card (NIC), Intel Corporation Wireless 8260, operating in promiscuous mode to monitor and collect the Wi-Fi network traffic.
For each UAV type, we collect the UAV traffic while they are flying and streaming video to the controller.
To do so, we set the channel frequency of the monitoring sensor in the same channel as the UAV's operating channel, then run Wireshark version 2.4.11 
to capture the Wi-Fi traffic data.
Each UAV type dataset contains 3,000 traffic traces with each trace having $n=200$ consecutive packets.

\begin{figure}[t] 
		  \centering
	 	\includegraphics[scale=0.75]{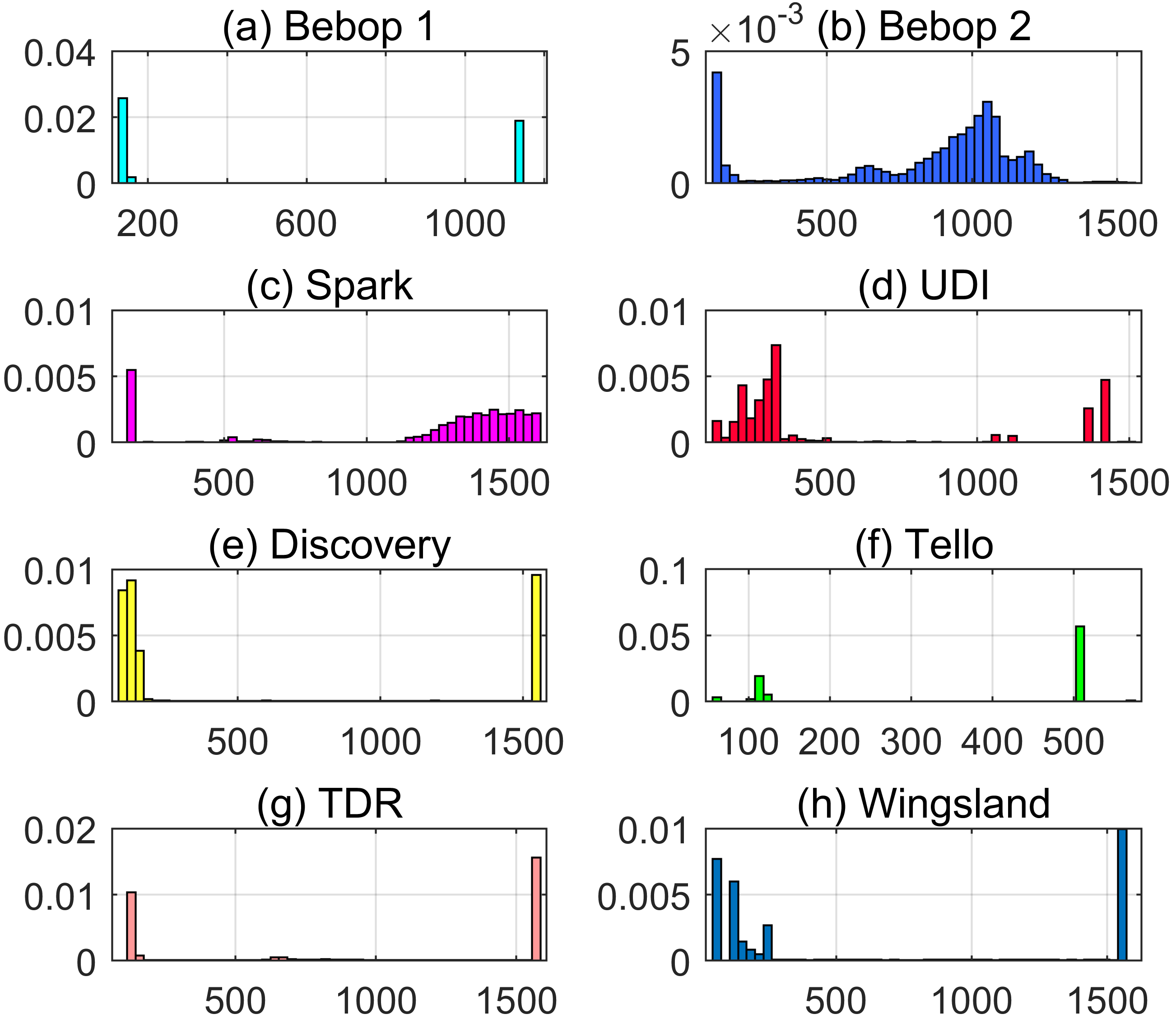} 
				\caption{Packet size distribution of different UAVs: $x$ and $y$ axes denote packet size and pdf, respectively.}
				\squeezeup
       \label{packetsizedist}
 \end{figure}

After collecting the data and identifying the UAVs' traffic flows, we clean the data and prepare it for the training and testing dataset.
In the data cleaning phase, we remove all the  broadcast packets (e.g., 802.11 beacon frames), damaged packets and packets with only receiving address (e.g., 802.11 ACK frames).
The remaining packets include video streaming, control commands, UAV's response to the control commands, and UAV status updates such as direction, velocity, height and GPS information.
In Fig. \ref{packetsizedist}, we show the packet size distribution of the UAV types used in our experiment.

Note that the data cleaning is performed in order to train and test the classification model offline.
However, when testing in the real scenario the incoming traffic may consist of broadcast packets, or packets with only ACK frames as well which all are easily discarded by the predefined filtering option on the packet capturing/monitoring software (i.e., Wireshark) before entering to the delay-aware UAV detection system.

 \begin{figure*}
  \subfigure[Training and testing accuracy on each subset $j$.]{\includegraphics[scale=0.755]{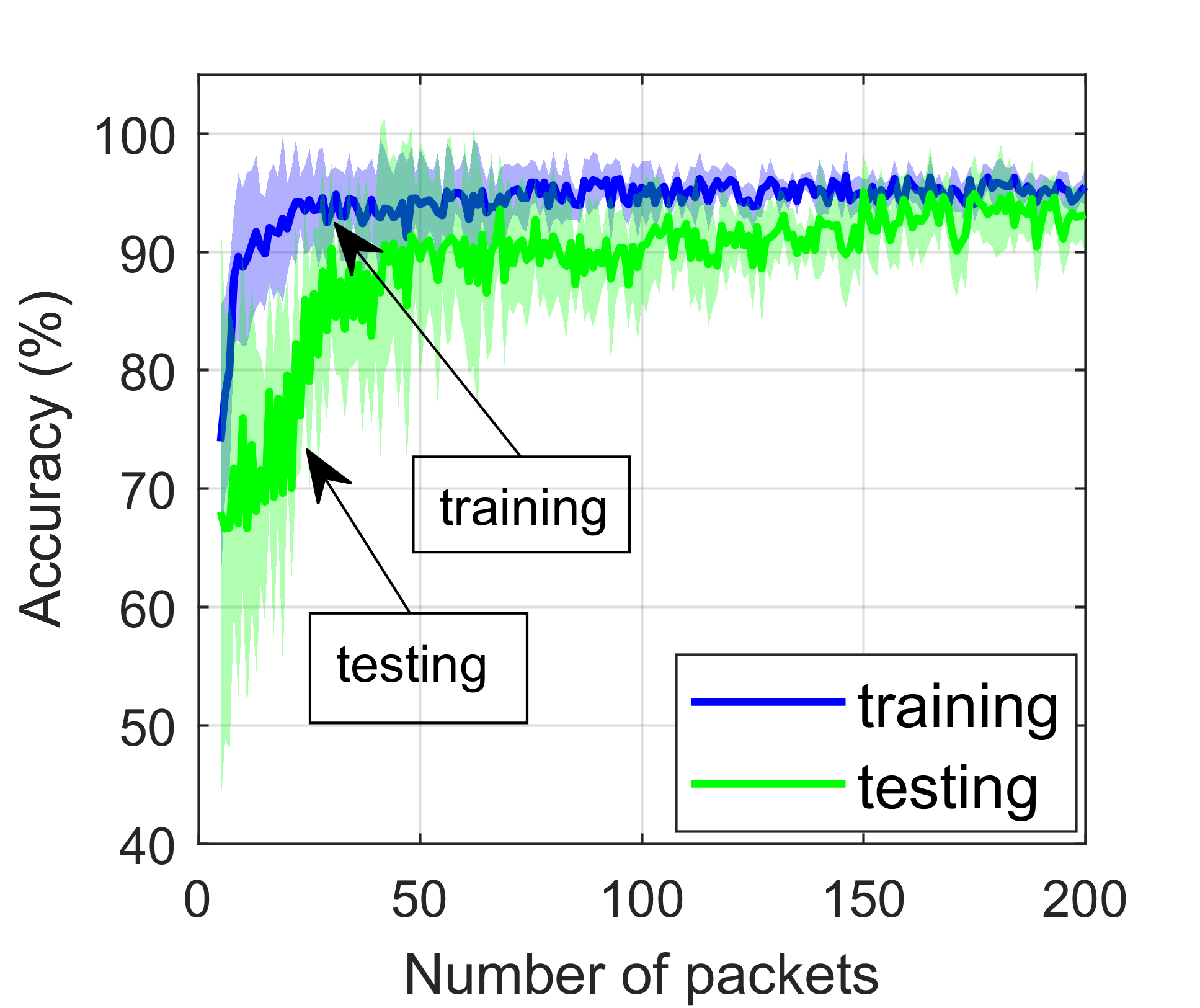}}
  \subfigure[F-measure metric.]{\includegraphics[scale=0.755]{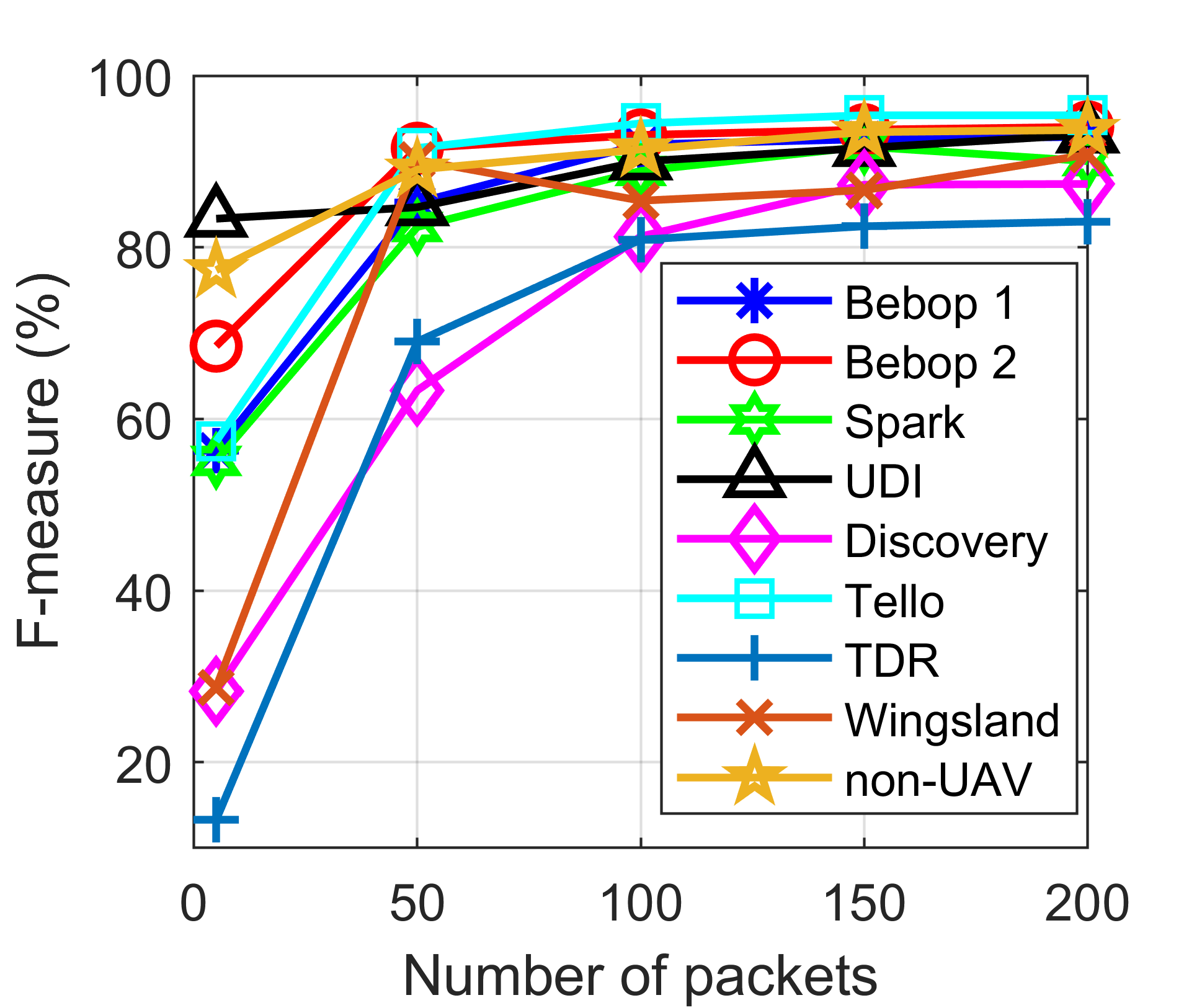}}
	  \subfigure[Linear discriminant analysis for $n=200$.]{\includegraphics[scale=0.445]{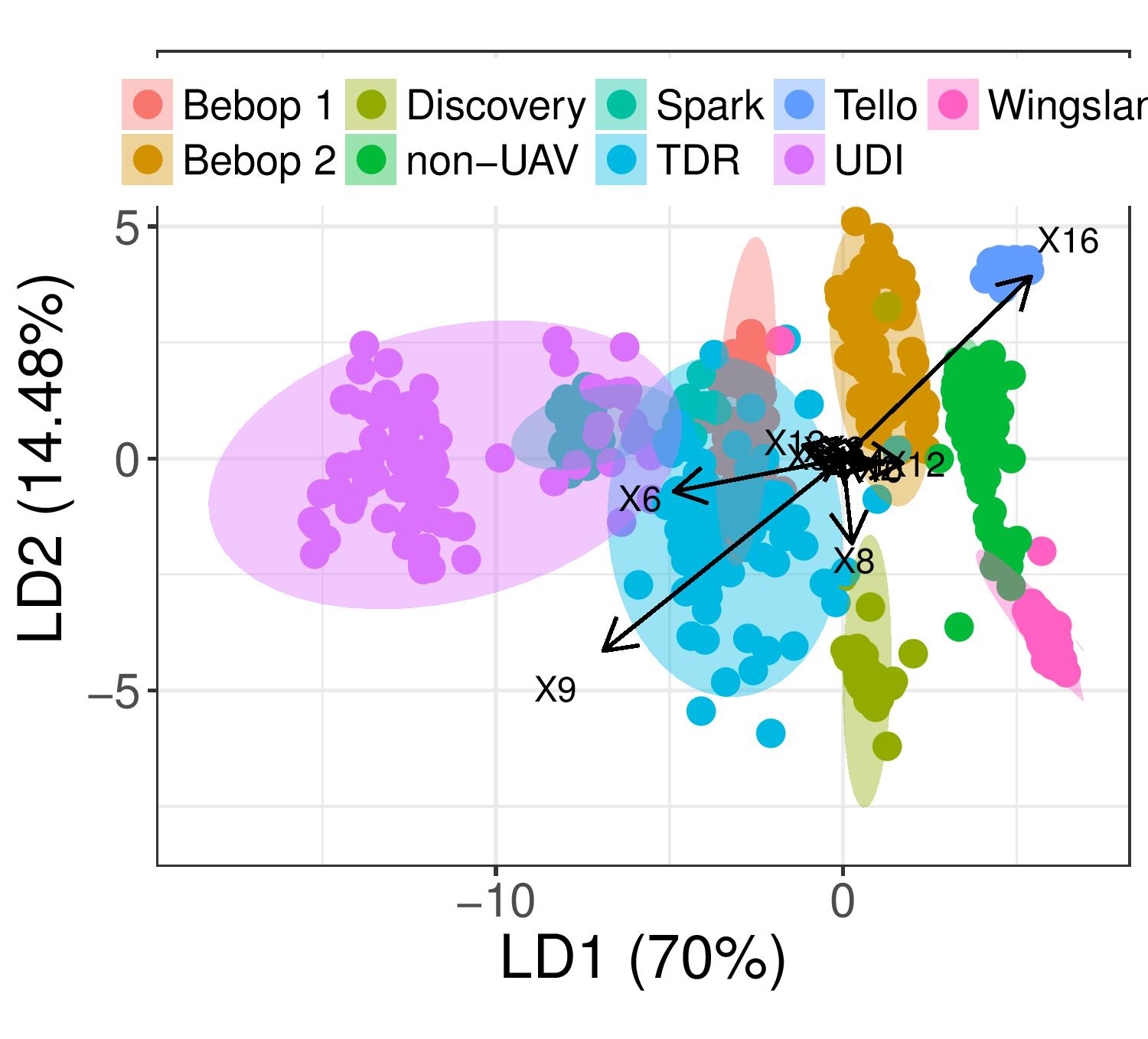}}
		\caption{Classification performance evaluation.}
				\squeezeup
				\label{traintestfig}
\end{figure*}
\begin{figure}[t] 
		  \centering
	 	\includegraphics[scale=0.16]{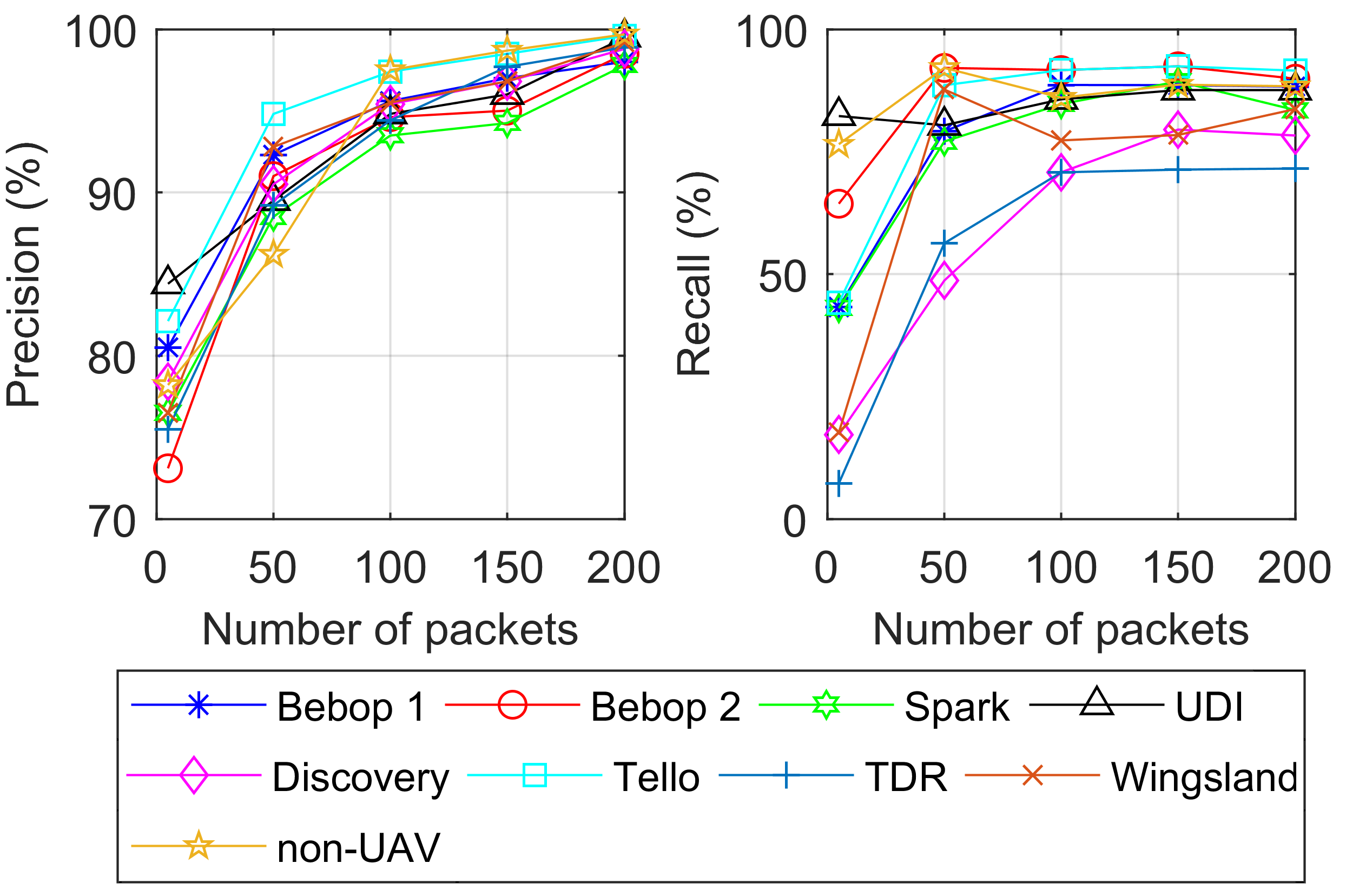} 
				\caption{Precision and recall metrics}.
       \label{jPR}
 \end{figure}

\subsection{Non-UAV Dataset}
In an effort to make a diverse non-UAV dataset, we create a dataset which consists of two main sub-dataset:
First, we use the Wi-Fi data traffic available online from CRAWDAD database \cite{CRAWDAD}.
We choose this dataset because of the following reasons.
1) This dataset consists of live and non-live video streaming traffic captured from commonly seen popular applications such as Google Hangouts, ooVoo, Skype, TED and Youtube.
2) The traffic data are collected from a smartphone app where the user makes a diverse set of mobility patterns.
Second, 
we have also captured encrypted Wi-Fi traffic on a university campus Wi-Fi network where a mixed multiple traffic types such as video streaming, social network apps, VoIP, email, web browsing applications are usually running.
If the UAV identification system is set up on the campus, our method should be able to differentiate UAV traffic from these non-UAV traffic.
The non-UAV dataset (Google Hangouts, ooVoo, Skype, TED, Youtube, and Campus traffic) also contains 3,000 traffic traces with $n=200$ consecutive packets.

\subsection{UAV Operation Mode Dataset}

The following steps are taken for an operation mode data traffic collection of a specific UAV type: 
1) Wi-Fi connection is established between the UAV and controller. 
2) A specific operation mode command (e.g., ``Forward'') is given via controller to the UAV and is held.
3) Wi-Fi medium monitoring sensor is activated to monitor the wireless channel traffic.
4) Wireshark is run on the promiscuous mode to capture the packets.
5) Before releasing the command in the controller, first, Wireshark is stopped, and then the collected traffic is saved and labeled according to the commanded operation mode.
7) This process is repeated for all the operation modes until enough data traffic is collected.

\section{Performance Evaluation}
\label{UAVDetectionImplementation}

\subsection{Learning-based Model Performance Evaluation}
By randomly sampling the dataset, we split the whole dataset into training and testing datasets with the ratio of $70\%$ and $30\%$, respectively. 
We create subset $S^j$ for $j=5,...,200$ by adding packet-by-packet information to each subset $j$ according to the step 2 in Algorithm 1.
Then, we form the design matrix $\bold{X}^j$ for $j=5,...,200$ by extracting $2l=24$ statistical feature values listed in Table~\ref{tab:feature}.
One-vs-all logistic regression multiclass classification algorithm with re-weighted $\ell_1$-norm technique proposed in (\ref{eq:objfun2}) is run over each design matrix $\bold{X}^j$.

Fig.~\ref{traintestfig}(a) illustrates the accuracy of the classification algorithm in training and testing on each design matrix $\bold{X}^j$. 
The shaded areas denote the regions surrounded by one standard deviation above and below the mean accuracy.
The results show that a mean testing accuracy of higher than 88\% is achieved when the information of fifty or more packets ($j>50$) are available in the subset. 
The model learning process also confirms the intuition that as more consecutive packets are available in the subset, the mean accuracy of predictive model increases. F-measure (weighted harmonic mean of precision and recall) for different UAV types and non-UAV is shown in Fig.~\ref{traintestfig}(b). Average F-measure of higher than 86\% is achieved on the test data for $n=200$. Fig.~\ref{jPR} also shows the corresponding precision and recall which indicate an acceptable discriminative power of the trained classifiers.

 \begin{figure*}
  \subfigure[ Selected (yellow) and discarded (blue) features.]{\includegraphics[scale=0.66]{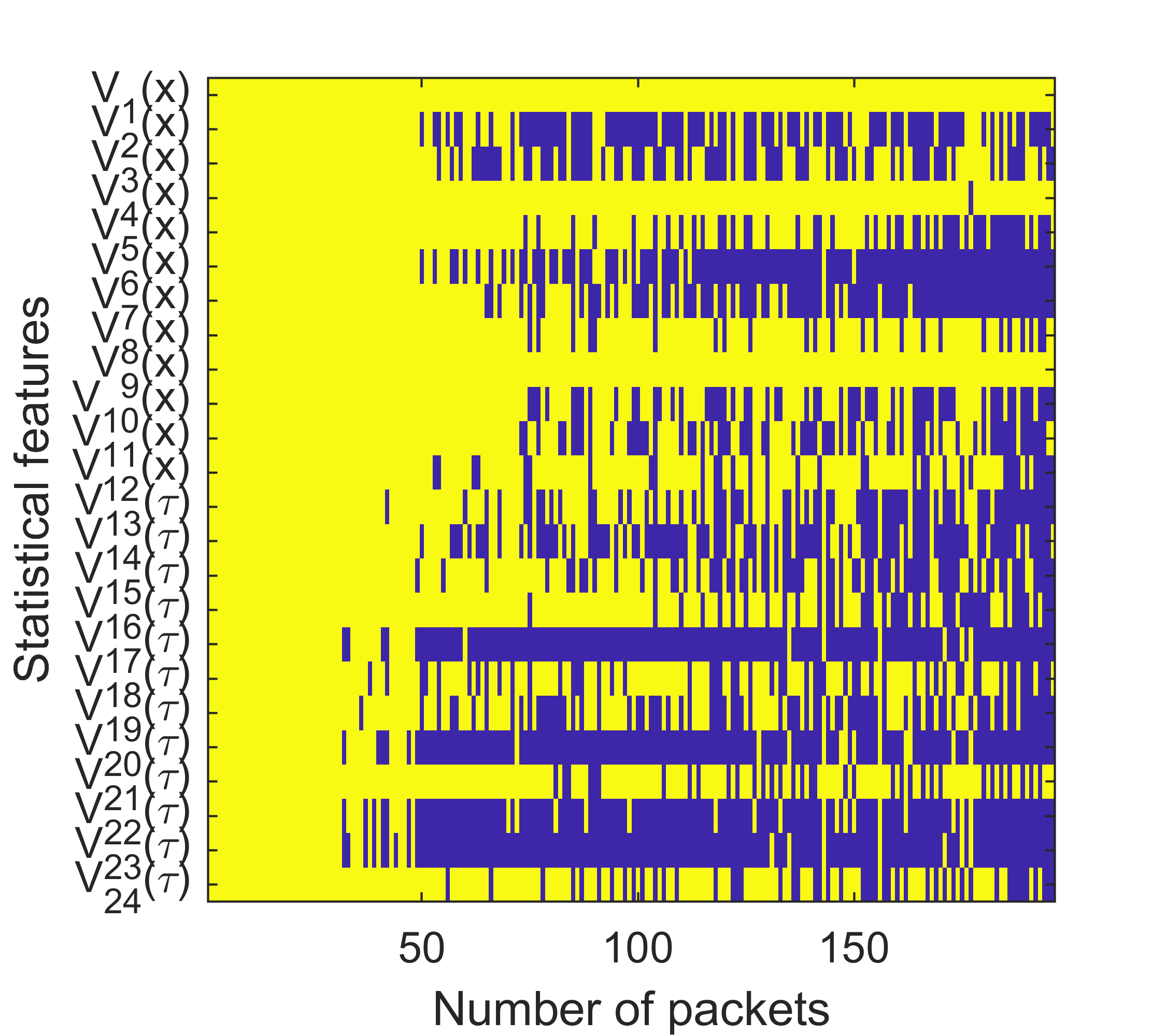}}
  \subfigure[Total feature computation time features.]{\includegraphics[scale=0.38]{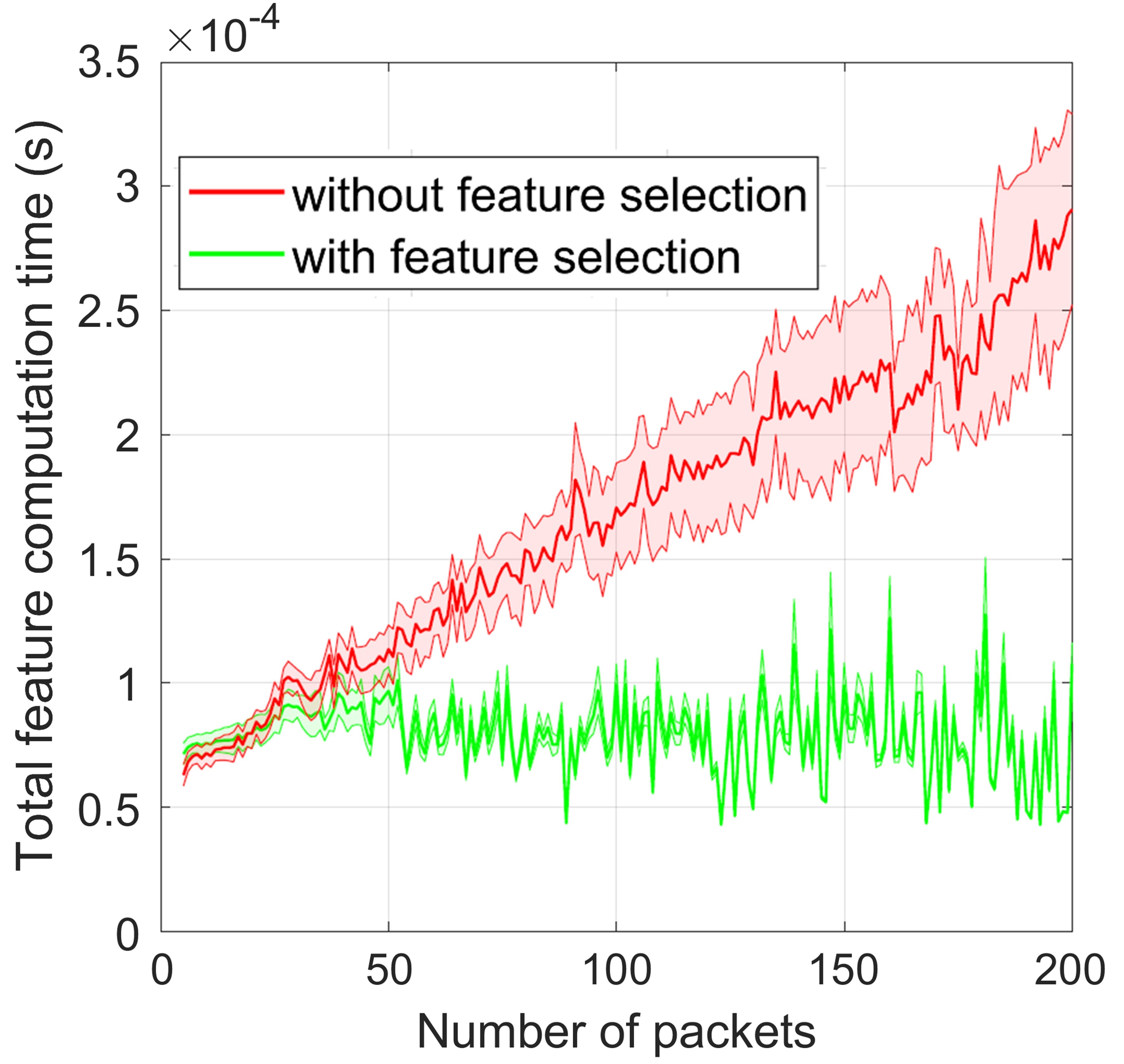}}
	  \subfigure[MSE. ]{\includegraphics[scale=0.66]{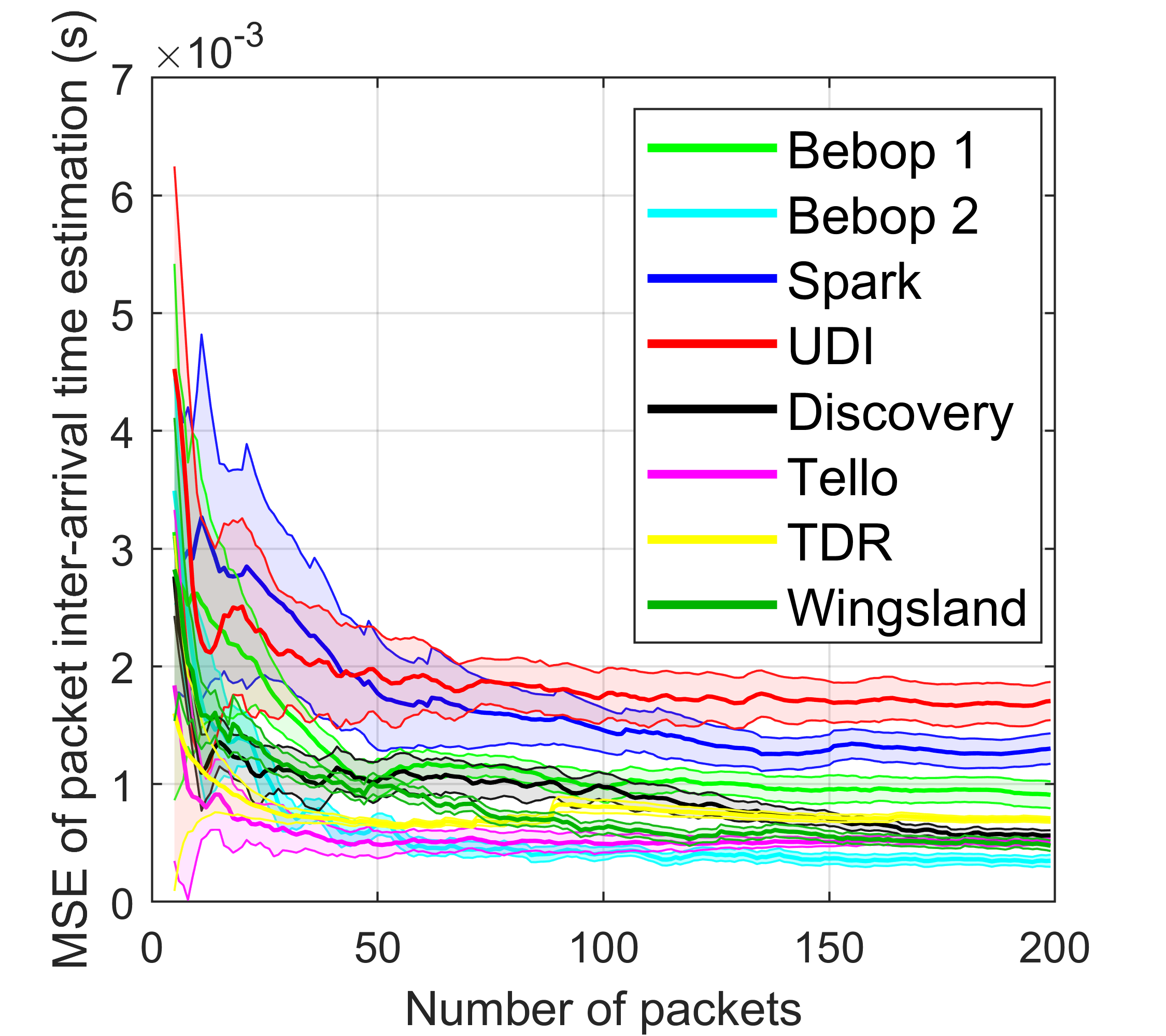}}
		\caption{Feature selection and packet inter-arrival time estimation performance evaluation.}
		\squeezeup
		\label{SelectedFeatures}
\end{figure*}


As a closely similar and related multiclass classification algorithm to one-vs-all logistic regression, we apply linear discriminant analysis (LDA) statistical method on the dataset \cite{murphy2013machine}.
This method is a generalized version of statistical Fisher's LDA. 
The LDA method finds a linear combination of the features to distinguish different classes in the dataset.
The output of this analysis is shown in Fig.~\ref{traintestfig}(c) for $n=200$.  
Various types of linear feature combination for different classes are shown in this figure with arrows followed by its associated feature index ($\Cross$1, $\Cross$2, ... , $\Cross$24). 
Successive discriminant function in the LDA analysis provides four proportions $LD1=0.70, LD2=0.1448, LD3=0.0480$, and $LD4=0.0232${\color{blue},} which describes the proportion of between-class variances. 
It is well visualized in this figure that how UAV types and non-UAVs are distinguished on $LD2$ verse $LD1$ as a result of the features' linear combination.  

\begin{figure*}
  \includegraphics[scale=.54]{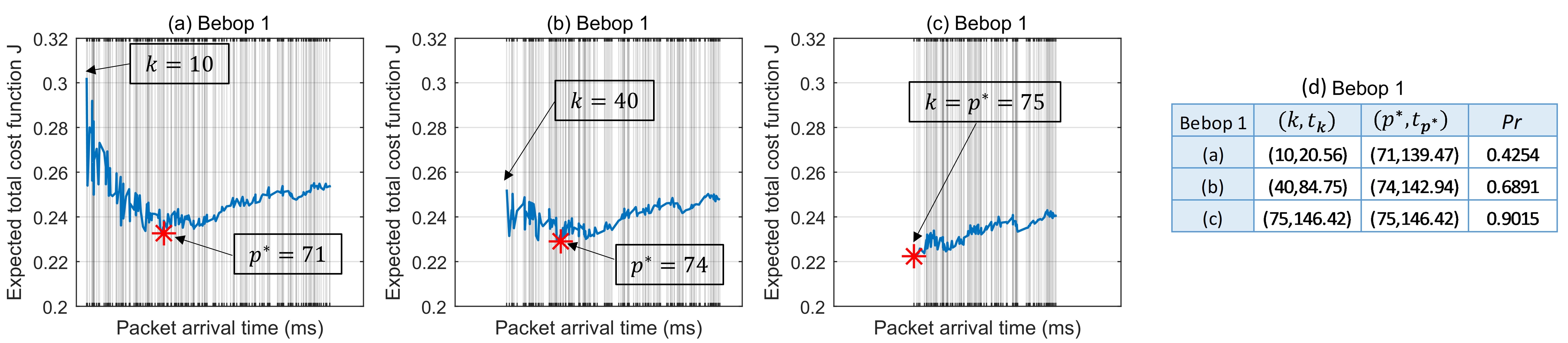}
		\caption{Delay-aware UAV identification using Algorithm 2.}
						 \squeezeup
        \label{UAVddfig}
\end{figure*} 

\subsection{Feature Selection and Computation Time} 
Our objective function in (\ref{eq:objfun2}) jointly minimizes the missclassification error and runtime by discarding useless features.
Fig.~\ref{SelectedFeatures}(a) shows the set of selected features for each model trained on $j$th subset for $j=5,...,200$.
As it is shown in the figure, when the sample size is small (e.g., $n<30$), all the features are selected by the model. 
This is because, on the one hand, small sample size does not provide enough information to the classifier to distinguish different classes with high accuracy, and on the other hand, it consumes less amount of time to compute the feature values. 
However, as the sample size increases, misclassification error is reduced and the feature computation time increases which results in the smaller set of selected features by the algorithm.

In order to evaluate the impact of feature selection method on the prediction time, we select 1,000 traces uniformly at random from the UAV dataset and consider them as incoming flows (i.e., $\bold{\tilde{x}}(t_k)$).
Then, we compute the feature generation time of the flows for $k=5,...,200$.
Fig.~\ref{SelectedFeatures}(b) illustrates the mean total feature generation time of the flows versus the number of the packets in the trace with and without feature selection method.
The shaded area denotes one standard deviation above and below the mean total computation time.
The results show that as the number of the packets increases, with feature selection, the mean total feature generation time oscillates in a non-increasing trend depending on the number of selected features.
However, without feature selection, the total computation time increases when the number of packets increases.
Therefore, the proposed feature selection method reduces the prediction runtime despite the fact that the sample size is increasing.  

\subsection{MLE Performance Evaluation}
For each UAV type, we select $1,000$ traces uniformly at random from the UAV dataset.
We consider the selected traces as incoming traffic flows.
Then, we follow the step 2 in Algorithm 2 to estimate the packet inter-arrival time of each flow.
Using (\ref{equ121}), we evaluate the MLE-based estimation performance.  
Fig. ~\ref{SelectedFeatures}(c) shows the mean MSE between the true and estimated packet inter-arrival time with shaded area of one standard deviation.
The results show that as more packets arrive, the mean MSE decreases. 
This means that as more packets are captured, the cost function $C_2$ estimation improves as the estimation accuracy of inter-arrival time enhances.
This results in achieving high quality delay-aware UAV identification.

\subsection{Delay-aware UAV Identification Test}

We consider eight types of incoming UAV traffic flows each belonging to a specific UAV type, and run the delay-aware UAV detection algorithm on them.
Due to the space limitation, we only show the test results for Bebop 1 in three steps Fig.~\ref{UAVddfig}(a), (b), (c), and summarize the outcome in the far right table in Fig.~\ref{UAVddfig}(d).

In Fig.~\ref{UAVddfig}(a), $k=10$th packet arrives at time $t_k=20.56ms$ and based on the received traffic flow till then, total cost function $J$ is estimated. 
In this case, it is estimated that the minimum total cost function will occur when $p^*=71$th packet arrives at $t_{p^*}=139.41 ms$.
Therefore, the decision for the flow detection is deferred. 
The far right column of the table in Fig.~\ref{UAVddfig}(d) indicates that if the UAV identification is performed in $k=10$, then the detection probability will be $42.54\%$ ($Pr=0.4254$).
In Fig.~\ref{UAVddfig}(b), $k=40$th packet arrives at time $t_k=84.75 ms$.
In this case, the algorithm estimates that the minimum expected total cost function will occur when $p^*=74$th packet arrives at $t_{p^*}=142.94 ms$.
If the identification is performed in $k=40$, then with the probability of $Pr=0.6891$ the flow will be detected as a Bebop 1 traffic flow.
This process is continued until the arrival of the $k$th packet for which $k=p^*$. 
According to Fig.~\ref{UAVddfig}(c), this condition is satisfied when $k=75$th packet arrives at $t_k=146.42 ms$ for which $k = p^*=75$. 
In this case, UAV detection is performed and the detection probability is $Pr=0.9015$.

Next, we select 1,000 traffic traces uniformly at random from the test dataset and test the flows based on the proposed delay-aware UAV early detection algorithm.
Table~\ref{tab:UAVflowtest} shows the test results where $E[p^*]=\frac{1}{N_\gamma} \sum_{i=1}^{N_\gamma} p_i^*$ denotes the average optimal number of packets, $E[t_{p^*}]=\frac{1}{N_\gamma} \sum_{i=1}^{N_\gamma} t_{p_i^*}$ indicates the average arrival time of $p^*$th packet where
$N_\gamma$ is the number of selected traces for class $\gamma$ and  $\gamma \in$ {\small $\left\{\text{Bebop 1},\text{Bebop 2},\text{Spark},\text{UDI},\text{Discovery},\text{Tello},\text{TDR},\text{Wingsland}\right\}$.}
In Table~\ref{tab:UAVflowtest}, accuracy is defined as the number of correct detection divided by the total number of traces selected for the test.
The results show that for the eight tested UAV types, our proposed method can detect and identify the UAVs in average within $0.15-0.35s$ with high average accuracy of $85.7-95.2\%$.

\begingroup
\setlength{\tabcolsep}{7pt} 
\renewcommand{\arraystretch}{1.1} 
\begin{table}
\small
\centering
\begin{threeparttable}
\caption{Tested UAVs' identification performance.\vspace{-0.1cm}}
\begin{tabular}{c|c|c|c}\hline\hline
Traffic & $E[p^*]$  & $E[t_{p^*}] (ms)$ &Accuracy (\%)\\\hline\hline 
Bebop 1&87 ($\pm$8)&160.43 ($\pm$10.01)&87.84 ($\pm$1.20)\\\hline
Bebop 2&95 ($\pm$13) &151.91 ($\pm$18.82)&90.75 ($\pm$1.74)\\\hline
Spark    &93 ($\pm$11)&142.80 ($\pm$15.57)&95.23 ($\pm$0.69)\\\hline
UDI     &141 ($\pm$21) &350.79 ($\pm$23.41)&85.76 ($\pm$2.38)\\\hline
Discovery& 94 ($\pm$3)  & 131.11 ($\pm$10.42)  & 92.52 ($\pm$0.85) \\\hline
Tello   & 72 ($\pm$7) & 121.65 ($\pm$31.30)  & 93.68 ($\pm$1.01)  \\\hline
TDR    & 68 ($\pm$13) & 100.77 ($\pm$12.75)  & 89.66 ($\pm$2.10)  \\\hline
Wingsland    &  75 ($\pm$18) &  92.46 ($\pm$ 21.22) & 94.39 ($\pm$1.85)   \\\hline
\hline
\end{tabular} 
\squeezeup
\label{tab:UAVflowtest}
\end{threeparttable}
\end{table}\normalsize
\endgroup

\subsection{UAV Detection Distance}
\begin{figure}[t] 
		  \centering
	 	\includegraphics[scale=0.42]{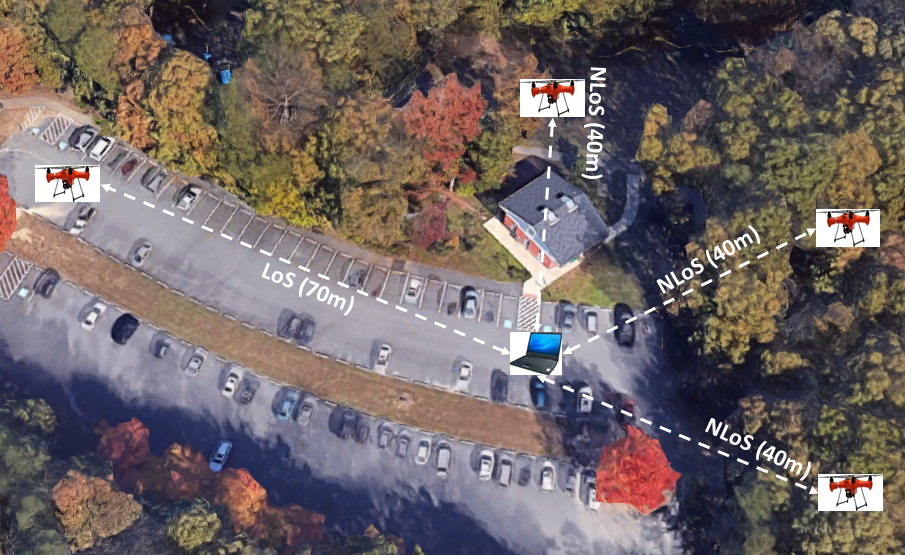} 
				\caption{UAV detection test scenarios}.
       \label{experiment}
 \end{figure}

UAV detection range is quite dependent on the Wi-Fi traffic monitoring sensor’s hardware specification (i.e., antenna type and gain). In this experiment, we have used a DELL Latitude laptop embedded with a wireless network interface card (NIC), Intel Corporation Wireless 8260, operating in promiscuous mode to monitor and collect the Wi-Fi network traffic. 
Considering this type of packet capturing system, for our experiment shown in Fig.~\ref{experiment}, in the line-of-sight (LoS) and non-line-of-sight (NLoS) i.e., blocked by a wall/trees, the system can detect the introducing UAV in the range of 70m and 40m, respectively. 
For the distances beyond these ranges due to heavy packet loss the detection accuracy reduces significantly. 
This will be our next challenging problem to tackle the UAV detection using traffic identification when the traffic suffers from packet loss.

\subsection{UAV Operation Mode Identification Evaluation}
\label{UAVOperationModeIdentification}

Consumer UAVs' operation mode capabilities maybe different from each other depending on the vendor specifications and manufacturing model. 
Here, for the UAV types, Bebop 1, Bebop 2 and DJI we identify eight common and popular operation modes as $ \Zb = $\{``Standby'', ``Hover'', ``Forward'', ``Backward'', ``Up'',  ``Down'', ``Right'', ``Left''\}.
However, FPV does not support the ``Hover'' mode, so we exclude this mode from set $\Zb$ for this type of UAV.

In order to identify the operation mode of these UAVs, we apply SVM and RF multiclass classifiers on the collected real-world data traffic.
For each UAV type, we train the SVM and RF predictive model packet-by-packet for $n=10,...,300$ by tuning the best model parameters for each subset.
For the SVM classification method, we utilize radial basis function (RBF) kernel and tune the best model parameters. 
For Bebop $1$, Bebop $2$, DJI, and FPV the total number of operation mode traffic traces in the training dataset is equal to $9600$, $9600$, $9600$, and $8400$, respectively.
By randomly sampling each dataset, we split the whole dataset into training, cross validation and test datasets with the ratio of $60\%$, $20\%$ and $20\%$, respectively.
Using $10$-fold cross validation repeated three times the best model parameters are tuned.
For example, for Bebop 1's operation mode identification when $n=300$, the best model tuned parameters are $C=64$ and $\epsilon = 0.15$ with the number of support vector machines of $\{173,139,142,131,126,174,150,142\}$ for each operation mode in set $\Zb$, respectively.

\begin{figure*}
  \includegraphics[width=18cm]{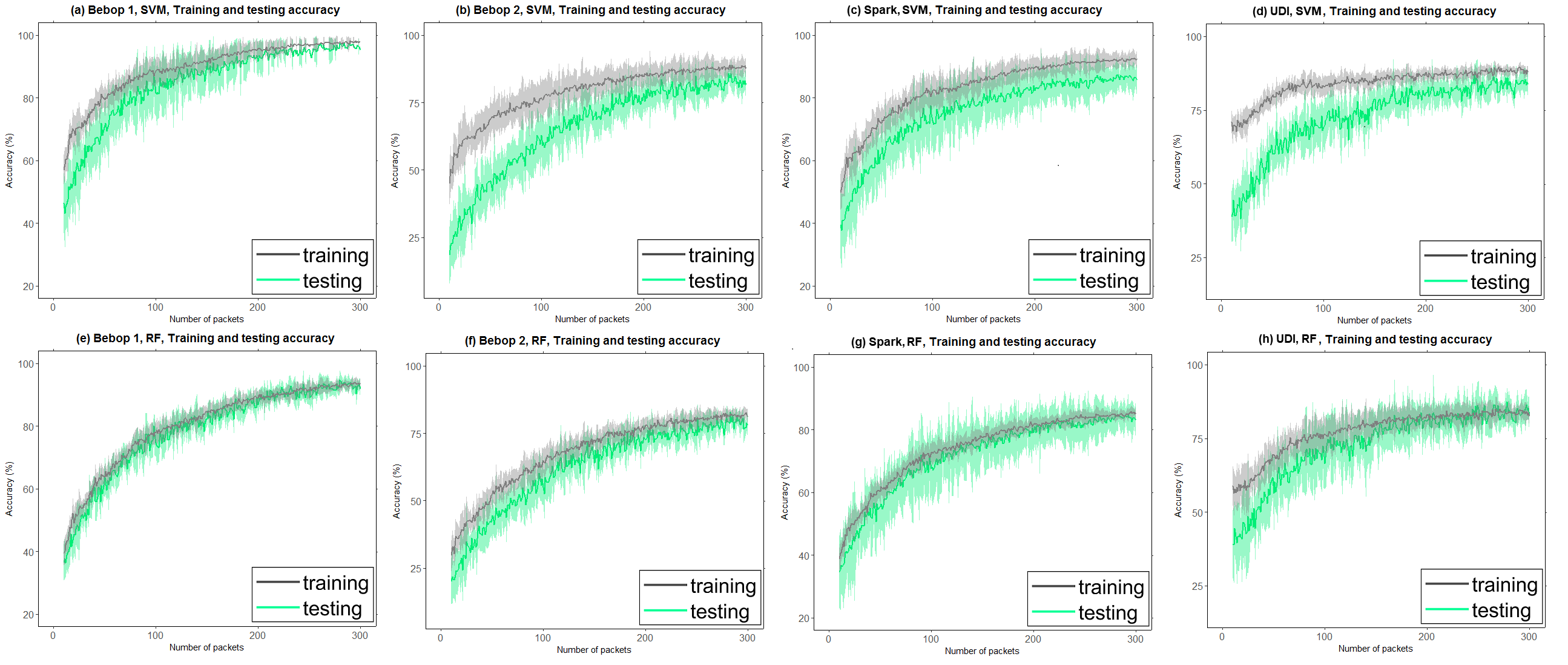}
		\caption{Training and testing accuracy of operation mode identification using SVM and RF classification algorithms.}
        \label{operationfig}
\end{figure*} 

\begin{figure*}
  \includegraphics[scale=.68]{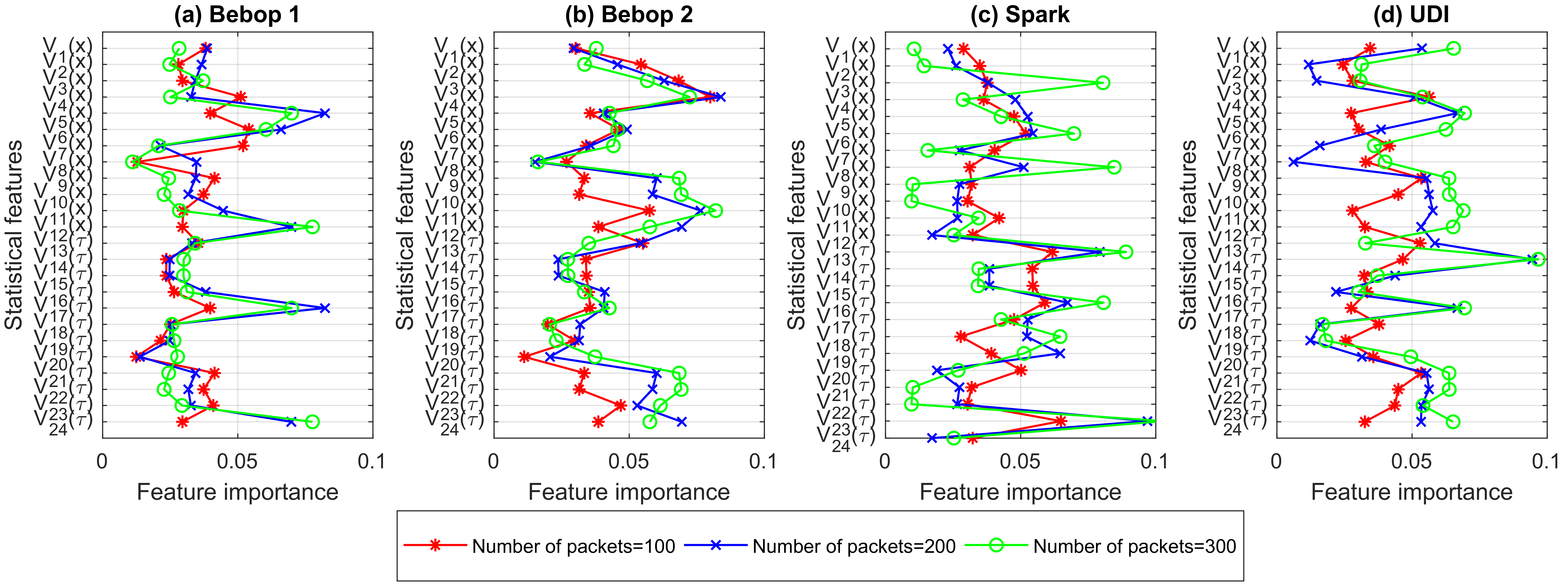}
		\caption{Feature importance analysis.}
        \label{FeatureImportance}
\end{figure*} 


\begin{figure*}
\centering
  \includegraphics[width=18cm]{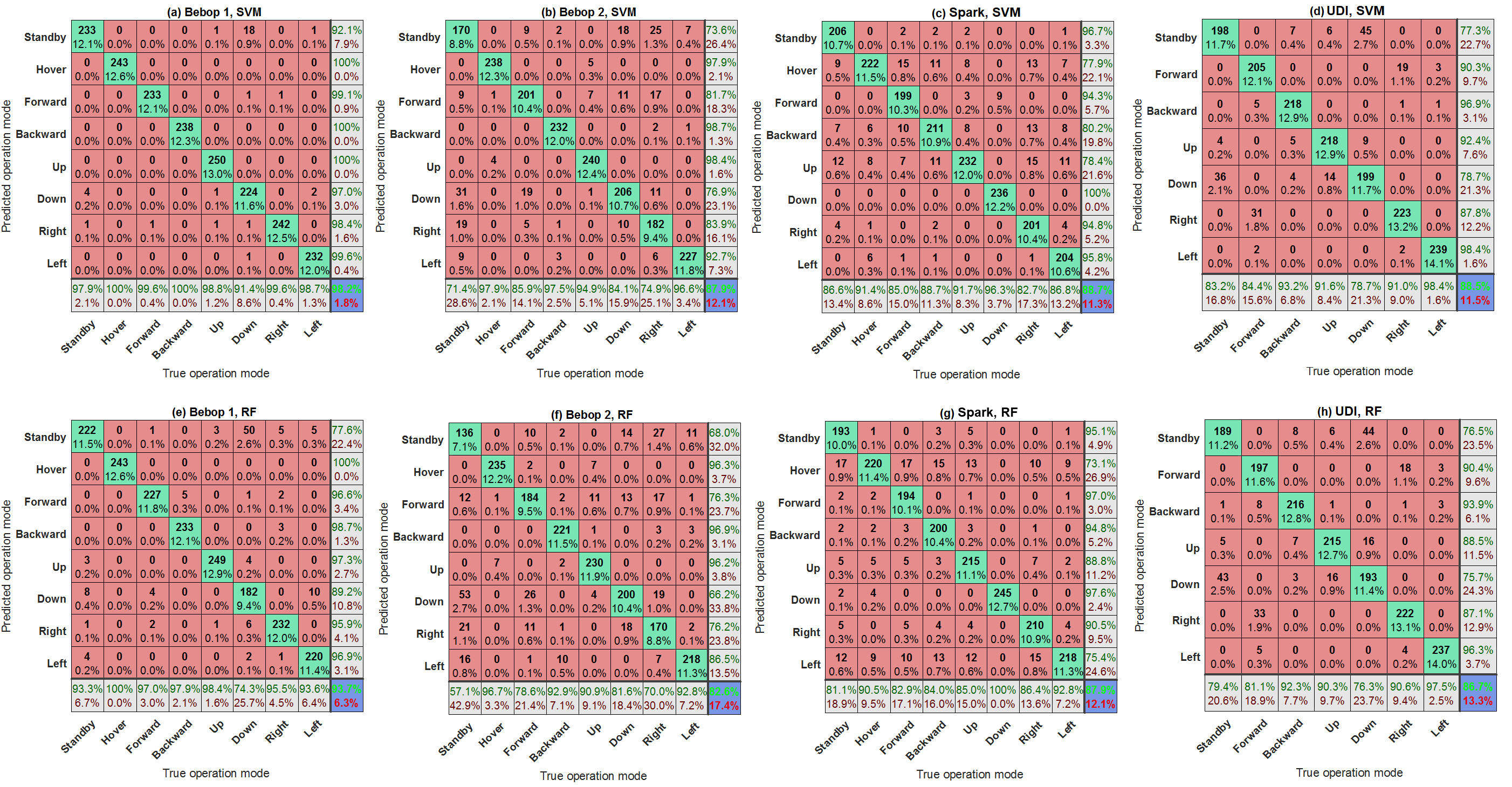}
		\caption{Operation mode identification confusion matrix, precision, recall and accuracy for different UAV types for $n=300$.}
        \label{CM}
\end{figure*}

Fig.~\ref{operationfig} illustrates the accuracy of the classification in training and testing for the four UAV types when SVM and RF are utilized for operation mode identification.
The gray (darker) and green (lighter) lines denote the mean accuracy of training and testing with the shaded area of one standard deviation, respectively.
Since the SVM and RF models are trained and cross validated for parameter tuning on different UAV types operation mode dataset, training and testing accuracy varies to some extent for each UAV type. However, both SVM and RF methods show an acceptable accuracy in effectively distinguishing the tested UAVs' operation modes.

Fig. \ref{FeatureImportance} illustrates the results of the feature importance analysis for various number of packets in the set. 
We can see that for different UAVs, the most important feature sets can be different.
For example, for Bebop 1 operation mode identification, 
$V_5(x)$,
$V_{12}(x)$ (skewness and  MAD of packet size),
$V_{17}(\tau)$, and
$V_{24} (\tau)$ (skewness and MAD of packet inter-arrival time) are indicating high importance value. 
However, for Bebop 2, 
$V_4(x)$,
$V_{11}(x)$ (STD and  PS of packet size),
$V_{21}(\tau)$,
$V_{22} (\tau)$,
$V_{24} (\tau)$ (Mean Square, RMS and MAD of packet inter-arrival time) are showing high importance value.
This indicates that: 1) For any given UAV type the data traffic patterns for various operation modes are different. 
2) The operation modes of each UAV type follows a different data traffic pattern than the other UAV types.
3) In order to train an effective model for the UAV operation mode identification, it is safe to consider all of statistical features so that the model can freely choose an effective set of important features which provides higher discriminative power.

Confusion matrices for operation mode identification of four tested UAVs when $n=300$ is shown in Fig.~\ref{CM}.
This figure indicates the overall performance of the SVM and RF multiclass classification algorithms.
In each confusion matrix, the diagonal and off-diagonal cells correspond to operation modes that are correctly and incorrectly identified, respectively. 
The right most column of the matrix indicates the precision (positive predictive value) and false discovery rate, as the top and bottom values of each cell, respectively.
Similarly, the bottom row of the matrix shows the recall (true positive rate) and false negative rate, in top and bottom part of each cell, respectively. 
Lastly, the cell in the most bottom right of the matrix, indicates the overall operation modes identification accuracy and error, respectively.
As the results show, the operation modes of the UAVs can be accurately identified with high accuracy of $88.5-98.2\%$ through wireless traffic fingerprinting.

\section{Discussion}
\label{Discussion}

\subsection{Significance of UAV Early Detection}
Considering that a consumer UAV can fly at 50-70mph and some racing UAVs could even fly above 150mph, a delay of one second will translate to a flying 
distance of 22m to 66m, which can be significant in practice for incident responses and safety/privacy protection.  
Therefore, reducing the detection delay is paramount important in the UAV invasion detection application.
Another improvement in detection time can be achieved using features' computational dependencies property  \cite{LZhao2018, Q2019}. 
Statistical features shown in Table \ref{tab:feature} are computationally dependent, so new techniques proposed in \cite{LZhao2018} can be applied to reduce the detection time. 
If number of the features are large, then large-scale feature computational dependency graph proposed in \cite{Q2019} can be employed to reduce the detection time as much as possible. 

\subsection{Applicability to Other Communication Protocols}
This work is based on the observation that many of consumer UAVs utilize Wi-Fi communication protocol for remote pilot control and video streaming. 
However, some type of consumer UAVs (e.g., DJI Phantom, 3DR Solo, Yuneec) may use other types of custom communication protocols such as Lightbridge, Sololink and Yuneec protocol.    
Our proposed framework is applicable to other types of consumer UAVs that use different communication protocols. 
Our framework works for encrypted wireless traffic and only needs packet size and inter-arrival time information (no need to use any packet content information). 
As long as we can obtain this information, our framework can be applied to not only detect the UAV using the proposed delay-aware mechanism, but also identify its operation mode.
 
In this paper, the hypothesis is that UAVs present unique traffic patterns that can be separated from other non-UAV traffic due to their use of a different set of communication protocols and physical operation.
We believe smart IP camera and handheld smartphone gimbal using a different set of communication protocols will be separable from UAV traffic as well.

\subsection{Recognizing New Types of UAVs}

In this paper, we applied supervised learning frameworks which can  classify the known classes (UAVs) appeared in the training set.
It will be interesting to extend our work to recognize new types of UAVs (unseen classes).
It belongs to the open set recognition problem which is still an open research problem in machine learning areas.
Existing technologies including \cite{e01RN299} could be explored to recognize new types of UAVs and at the same time reducing the model retraining overhead.
Although only a limited number of types of UAVs are tested in this work, the proposed framework should be able to handle a large dataset of different UAV subtypes.
Through experiments, this paper has demonstrated the discriminative power of the proposed classifier, which indicates the effectiveness of proposed methodologies.
The users are free to adjust dataset to cover different applications on different UAV subtypes.
Through experiments, this paper has demonstrated the discriminative power of the proposed classifier, which indicates the effectiveness of proposed methodologies.
The users are free to adjust both UAV and non-UAV dataset to cover different application scenarios (e.g., university campus, government building, airport, etc).

\subsection{More Sophisticated Scenarios}
The framework developed in this paper could be extended to tackle more sophisticated scenarios, such as simultaneous detection of UAVs operating on multiple channels.
A more powerful adversary could even hop among different channels to escape from detection.
Some multi-channel network monitoring mechanisms \cite{YXPZ2016} could be integrated in this scenario.
An intelligent adversary could change its traffic pattern by injecting  packets to avoid being detected.
However, this kind of adversary could be limited by energy budget (i.e., limited number of packets can be injected due to limited battery capacity) and mission requirement (i.e., genuine command control packets and video streaming packets cannot be suppressed).
Therefore, an enhanced machine learning model which can effectively test sub-traffic could still be effective when facing such an intelligent adversary.
Furthermore, investigating the possibility of combining traffic information and physical layer information (such as RSS and modulation schemes) to enhance the identification performance and enable UAV localization is of great interest.

\section{Conclusions}
\label{conclusion}
Detecting and identifying consumer UAVs is of utmost importance for regulation enforcement, forensics investigation, public security, and personal privacy protection. 
To complement existing physical detection mechanisms, we proposed a delay-aware machine learning-based UAV detection and operation mode identification framework over encrypted Wi-Fi UAV traffic.
This framework extracts features from packet size and inter-arrival time and in the model training phase adopts re-weighted $\ell_1$-norm regularization with consideration of computation time among various features.
Therefore, feature selection and performance optimization are integrated in one objective function.
To deal with packet inter-arrival time uncertainty when estimating the cost function, we utilized model-based MLE method to estimate the packet inter-arrival times of the incoming flow.
 We collected a large amount of encrypted Wi-Fi traffic of eight types of consumer UAVs and conducted extensive evaluation on the performance of our proposed methods.
Experimental results show that the proposed methods can detect and identify tested UAVs within $0.15-0.35s$ with the accuracy of $85.7-95.2\%$.
The UAV detection range is within the physical sensing range of 70m and 40m in the line-of-sight (LoS) and non-line-of-sight (NLoS) scenarios, respectively.
The operation modes of UAVs can also be well identified with accuracy in the range of $88.5-98.2\%$.
The operation mode identification reveals the cyber-physical coupling property of UAVs. 
Based on this coupling, we can infer information on the physical status (operation mode) of UAVs given information on their cyber part (Wi-Fi traffic data).

Although this work uses Wi-Fi traffic to detect and identify consumer UAVs, we believe the proposed machine learning-based detection framework and methodology are general enough to be applied to other cyber-physical/IoT systems using different wireless communication technologies (e.g., Bluetooth and cellular).
We hope this work to shed light on the cyber-physical attack co-detection or co-defense for many other CPS/IoT systems.

\bibliographystyle{IEEEtran}
\bibliography{ccssample}

\begin{thebibliography}{10}
\providecommand{\url}[1]{#1}
\csname url@samestyle\endcsname
\providecommand{\newblock}{\relax}
\providecommand{\bibinfo}[2]{#2}
\providecommand{\BIBentrySTDinterwordspacing}{\spaceskip=0pt\relax}
\providecommand{\BIBentryALTinterwordstretchfactor}{4}
\providecommand{\BIBentryALTinterwordspacing}{\spaceskip=\fontdimen2\font plus
\BIBentryALTinterwordstretchfactor\fontdimen3\font minus
  \fontdimen4\font\relax}
\providecommand{\BIBforeignlanguage}[2]{{%
\expandafter\ifx\csname l@#1\endcsname\relax
\typeout{** WARNING: IEEEtran.bst: No hyphenation pattern has been}%
\typeout{** loaded for the language `#1'. Using the pattern for}%
\typeout{** the default language instead.}%
\else
\language=\csname l@#1\endcsname
\fi
#2}}
\providecommand{\BIBdecl}{\relax}
\BIBdecl

\bibitem{ARYA2017}
\BIBentryALTinterwordspacing
R.~Altawy and A.~M. Youssef, ``Security, privacy, and safety aspects of
  civilian drones: A survey,'' \emph{ACM Trans. Cyber-Phys. Syst.}, vol.~1,
  no.~2, pp. 7:1--7:25, Nov. 2016. [Online]. Available:
  \url{http://doi.acm.org/10.1145/3001836}
\BIBentrySTDinterwordspacing

\bibitem{RT2017crash}
\BIBentryALTinterwordspacing
D.~Furfaro, L.~Celona, and N.~Musumeci, ``Civilian drone crashes into army
  helicopter,'' RT, 2017. [Online]. Available:
  \url{http://nypost.com/2017/09/22/army-helicopter-hit-by-drone/}
\BIBentrySTDinterwordspacing

\bibitem{RT2016peeping}
\BIBentryALTinterwordspacing
RT, ``Peeping drone: {UAV} hovers outside of massachusetts teen's bedroom
  window,'' RT, Apr. 2016. [Online]. Available:
  \url{https://www.rt.com/usa/341404-drone-privacy-teenager-window/}
\BIBentrySTDinterwordspacing

\bibitem{shear2015white}
\BIBentryALTinterwordspacing
M.~Shear and M.~Schmidt, ``White {H}ouse drone crash described as a {US}
  worker’s drunken lark,'' New York Times, Jan. 2015. [Online]. Available:
  \url{https://www.nytimes.com/2015/01/28/us/white-house-drone.html}
\BIBentrySTDinterwordspacing

\bibitem{FAA}
F.~A. Administration, ``{UAS} registration,'' FAA website:
  https://www.faa.gov/uas/getting\_started/registration/, FAA, 2015.

\bibitem{moses2011radar}
A.~Moses, M.~J. Rutherford, and K.~P. Valavanis, ``Radar-based detection and
  identification for miniature air vehicles,'' in \emph{Proc. IEEE
  International Conference on Control Applications (CCA)}, Sep. 2011, pp.
  933--940.

\bibitem{DH2017}
D.~H. Shin, D.~H. Jung, D.~C. Kim, J.~W. Ham, and S.~O. Park, ``A distributed
  fmcw radar system based on fiber-optic links for small drone detection,''
  \emph{IEEE Transactions on Instrumentation and Measurement}, vol.~66, no.~2,
  pp. 340--347, Feb 2017.

\bibitem{zelnio2009low}
A.~M. Zelnio, E.~E. Case, and B.~D. Rigling, ``A low-cost acoustic array for
  detecting and tracking small rc aircraft,'' in \emph{Digital Signal
  Processing Workshop and 5th IEEE Signal Processing Education Workshop, 2009.
  DSP/SPE 2009. IEEE 13th}, Jan 2009, pp. 121--125.

\bibitem{marmaroli2012uav}
P.~Marmaroli, X.~Falourd, and H.~Lissek, ``A {UAV} motor denoising technique to
  improve localization of surrounding noisy aircrafts: proof of concept for
  anti-collision systems,'' in \emph{Acoustics}, April 2012, pp. 23--27.

\bibitem{rozantsev2017detecting}
A.~Rozantsev, V.~Lepetit, and P.~Fua, ``Detecting flying objects using a single
  moving camera,'' \emph{in Proc. IEEE Transactions on Pattern Analysis and
  Machine Intelligence}, vol.~39, no.~5, pp. 879--892, May 2017.

\bibitem{PA2018}
P.~A. Prates, R.~Mendonça, A.~Lourenço, F.~Marques, J.~P. Matos-Carvalho, and
  J.~Barata, ``Vision-based {UAV} detection and tracking using motion
  signatures,'' in \emph{Proc. IEEE Industrial Cyber-Physical Systems (ICPS)},
  May 2018, pp. 482--487.

\bibitem{AR2015}
A.~Rozantsev, V.~Lepetit, and P.~Fua, ``Flying objects detection from a single
  moving camera,'' in \emph{2015 IEEE Conference on Computer Vision and Pattern
  Recognition (CVPR)}, June 2015, pp. 4128--4136.

\bibitem{NJ2014}
N.~Jing, M.~Yang, S.~Cheng, Q.~Dong, and H.~Xiong, ``An efficient {SVM}-based
  method for multi-class network traffic classification,'' in \emph{Proc. 30th
  IEEE International Performance Computing and Communications Conference}, Nov
  2011, pp. 1--8.

\bibitem{AMMH2004}
A.~McGregor, M.~Hall, P.~Lorier, and J.~Brunskill, ``Flow clustering using
  machine learning techniques,'' in \emph{Passive and Active Network
  Measurement}, C.~Barakat and I.~Pratt, Eds.\hskip 1em plus 0.5em minus
  0.4em\relax Berlin, Heidelberg: Springer Berlin Heidelberg, 2004, pp.
  205--214.

\bibitem{Bar2010}
R.~Bar~Yanai, M.~Langberg, D.~Peleg, and L.~Roditty, ``Realtime classification
  for encrypted traffic,'' in \emph{Experimental Algorithms}.\hskip 1em plus
  0.5em minus 0.4em\relax Springer, 2010, pp. 373--385.

\bibitem{AD2015}
A.~Dachraoui, A.~Bondu, and A.~Cornu{\'e}jols, ``Early classification of time
  series as a non myopic sequential decision making problem,'' in \emph{Machine
  Learning and Knowledge Discovery in Databases}.\hskip 1em plus 0.5em minus
  0.4em\relax Cham: Springer International Publishing, 2015, pp. 433--447.

\bibitem{MA2014}
A.~Moses, M.~J. Rutherford, M.~Kontitsis, and K.~P. Valavanis, ``{UAV}-borne
  {X}-band radar for collision avoidance,'' \emph{Robotica}, vol.~32, no.~1,
  pp. 97--114, 2014.

\bibitem{MAMR2011}
A.~Moses, M.~J. Rutherford, and K.~P. Valavanis, ``Radar-based detection and
  identification for miniature air vehicles,'' in \emph{Proc. IEEE
  International Conference on Control Applications (CCA)}, Sep. 2011, pp.
  933--940.

\bibitem{GMTR2016}
G.~J. Mendis, T.~Randeny, J.~Wei, and A.~Madanayake, ``Deep learning based
  doppler radar for micro uas detection and classification,'' in \emph{IEEE
  MILCOM 2016}, Nov 2016, pp. 924--929.

\bibitem{GF2015}
F.~G{\"o}kçe, G.~{\"U}çoluk, E.~Sahin, and S.~Kalkan, ``Vision-based
  detection and distance estimation of micro unmanned aerial vehicles,'' in
  \emph{Sensors}, September 2015.

\bibitem{sutin2013acoustic}
A.~Sutin, H.~Salloum, A.~Sedunov, and N.~Sedunov, ``Acoustic detection,
  tracking and classification of low flying aircraft,'' in \emph{Proc.
  Technologies for Homeland Security (HST)}, Nov 2013, pp. 141--146.

\bibitem{PMXF2012}
P.~Marmaroli, X.~Falourd, and H.~Lissek, ``A {UAV} motor denoising technique to
  improve localization of surrounding noisy aircrafts: proof of concept for
  anti-collision systems,'' in \emph{Acoustics}, April 2012, pp. 23--27.

\bibitem{JBFP2015}
J.~Busset, F.~Perrodin, P.~Wellig, B.~Ott, K.~Heutschi, T.~R{\"u}hl, and
  T.~Nussbaumer, ``Detection and tracking of drones using advanced acoustic
  cameras,'' in \emph{Unmanned/Unattended Sensors and Sensor Networks XI; and
  Advanced Free-Space Optical Communication Techniques and Applications}, vol.
  9647, Oct 2015.

\bibitem{WSGA2011}
W.~Shi, G.~Arabadjis, B.~Bishop, P.~Hill, R.~Plasse, and J.~Yoder, ``Detecting,
  tracking, and identifying airborne threats with netted sensor fence,'' in
  \emph{Sensor Fusion-Foundation and Applications}.\hskip 1em plus 0.5em minus
  0.4em\relax InTech, 2011.

\bibitem{Zhao2018}
C.~{Zhao}, C.~{Chen}, Z.~{Cai}, M.~{Shi}, X.~{Du}, and M.~{Guizani},
  ``Classification of small uavs based on auxiliary classifier wasserstein
  gans,'' in \emph{2018 IEEE Global Communications Conference (GLOBECOM)}, Dec
  2018, pp. 206--212.

\bibitem{Bis2018}
I.~{Bisio}, C.~{Garibotto}, F.~{Lavagetto}, A.~{Sciarrone}, and S.~{Zappatore},
  ``Unauthorized amateur uav detection based on wifi statistical fingerprint
  analysis,'' \emph{IEEE Communications Magazine}, vol.~56, no.~4, pp.
  106--111, April 2018.

\bibitem{Martin2019}
M.~Ezuma, F.~Erden, C.~K. Anjinappa, O.~Ozdemir, and I.~Guvenc, ``Micro-uav
  detection and classification from rf fingerprints using machine learning
  techniques,'' \emph{2019 IEEE Aerospace Conference}, pp. 1--13, Dec 2019.

\bibitem{SBRB2017}
S.~Birnbach, R.~Baker, and I.~Martinovic, ``Wi-{F}ly?: Detecting privacy
  invasion attacks by consumer drones,'' in \emph{Proc. 24th Annual Network and
  Distributed System Security Symposium, {NDSS}}, Feb 2017.

\bibitem{Nguyen2017}
P.~Nguyen, H.~Truong, M.~Ravindranathan, A.~Nguyen, R.~Han, and T.~Vu,
  ``Matthan: Drone presence detection by identifying physical signatures in the
  drone's rf communication,'' in \emph{Proceedings of the 15th Annual
  International Conference on Mobile Systems, Applications, and Services}, ser.
  MobiSys.\hskip 1em plus 0.5em minus 0.4em\relax New York, USA: ACM, 2017, pp.
  211--224.

\bibitem{Sci2019}
\BIBentryALTinterwordspacing
S.~Sciancalepore, O.~A. Ibrahim, G.~Oligeri, and R.~Di~Pietro, ``Detecting
  drones status via encrypted traffic analysis,'' in \emph{Proceedings of the
  ACM Workshop on Wireless Security and Machine Learning}, ser. WiseML
  2019.\hskip 1em plus 0.5em minus 0.4em\relax New York, NY, USA: ACM, 2019,
  pp. 67--72. [Online]. Available:
  \url{http://doi.acm.org/10.1145/3324921.3328791}
\BIBentrySTDinterwordspacing

\bibitem{GM2012}
G.~Xie, M.~Iliofotou, R.~Keralapura, M.~Faloutsos, and A.~Nucci, ``Subflow:
  Towards practical flow-level traffic classification,'' in \emph{Proc. IEEE
  INFOCOM}, March 2012, pp. 2541--2545.

\bibitem{TNGA2008}
T.~T. Nguyen and G.~Armitage, ``A survey of techniques for internet traffic
  classification using machine learning,'' \emph{IEEE Communications Surveys \&
  Tutorials}, vol.~10, no.~4, pp. 56--76, April 2008.

\bibitem{CPS-Sec2019}
A.~{Alipour-Fanid}, M.~{Dabaghchian}, N.~{Wang}, P.~{Wang}, L.~{Zhao}, and
  K.~{Zeng}, ``Machine learning-based delay-aware {UAV} detection over
  encrypted wi-fi traffic,'' in \emph{Proc. IEEE CNS 2019 - IEEE International
  Workshop on Cyber-Physical Systems Security (CPS-SEC)}, June 2019.

\bibitem{FBRJ2012}
F.~Bach, R.~Jenatton, J.~Mairal, G.~Obozinski \emph{et~al.}, ``Optimization
  with sparsity-inducing penalties,'' \emph{Foundations and Trends in Machine
  Learning}, vol.~4, no.~1, pp. 1--106, 2012.

\bibitem{murphy2013machine}
K.~P. Murphy, \emph{Machine learning : a probabilistic perspective}.\hskip 1em
  plus 0.5em minus 0.4em\relax Cambridge, Mass. [u.a.]: MIT Press, 2013.

\bibitem{AK2010}
A.~Komaee, ``Maximum likelihood and minimum mean squared error estimations for
  measurement of light intensity,'' in \emph{2010 44th Annual Conference on
  Information Sciences and Systems (CISS)}, March 2010.

\bibitem{CRAWDAD}
S.~Sengupta, H.~Gupta, N.~Ganguly, B.~Mitra, P.~De, and S.~Chakraborty,
  ``{CRAWDAD} dataset iitkgp/apptraffic (v. 2015-11-26),'' Downloaded from
  https://crawdad.org/iitkgp/apptraffic/20151126, Nov. 2015.

\bibitem{LZhao2018}
\BIBentryALTinterwordspacing
L.~Zhao, A.~Alipour-Fanid, M.~Slawski, and K.~Zeng, ``Prediction-time efficient
  classification using feature computational dependencies,'' in
  \emph{Proceedings of the 24th ACM SIGKDD International Conference on
  Knowledge Discovery \&\#38; Data Mining}, ser. KDD '18.\hskip 1em plus 0.5em
  minus 0.4em\relax New York, NY, USA: ACM, 2018, pp. 2787--2796. [Online].
  Available: \url{http://doi.acm.org/10.1145/3219819.3220117}
\BIBentrySTDinterwordspacing

\bibitem{Q2019}
Q.~{Li}, A.~{Alipour-Fanid}, M.~{Slawski}, Y.~{Ye}, L.~{Wu}, K.~{Zeng}, and
  L.~{Zhao}, ``Large-scale cost-aware classification using feature
  computational dependency graph,'' \emph{IEEE Transactions on Knowledge and
  Data Engineering}, pp. 1--1, 2019.

\bibitem{e01RN299}
X.~{Mu}, K.~M. {Ting}, and Z.~{Zhou}, ``Classification under streaming emerging
  new classes: A solution using completely-random trees,'' \emph{IEEE
  Transactions on Knowledge and Data Engineering}, vol.~29, no.~8, pp.
  1605--1618, Aug 2017.

\bibitem{YXPZ2016}
Y.~Xue, P.~Zhou, T.~Jiang, S.~Mao, and X.~Huang, ``Distributed learning for
  multi-channel selection in wireless network monitoring,'' in \emph{2016 13th
  Annual IEEE International Conference on Sensing, Communication, and
  Networking (SECON)}, June 2016, pp. 1--9.

\end{thebibliography}

\vskip 0pt plus -0.8fil

\begin{IEEEbiography}
[{\includegraphics[width=1.2in,height=1.3in,clip,keepaspectratio]{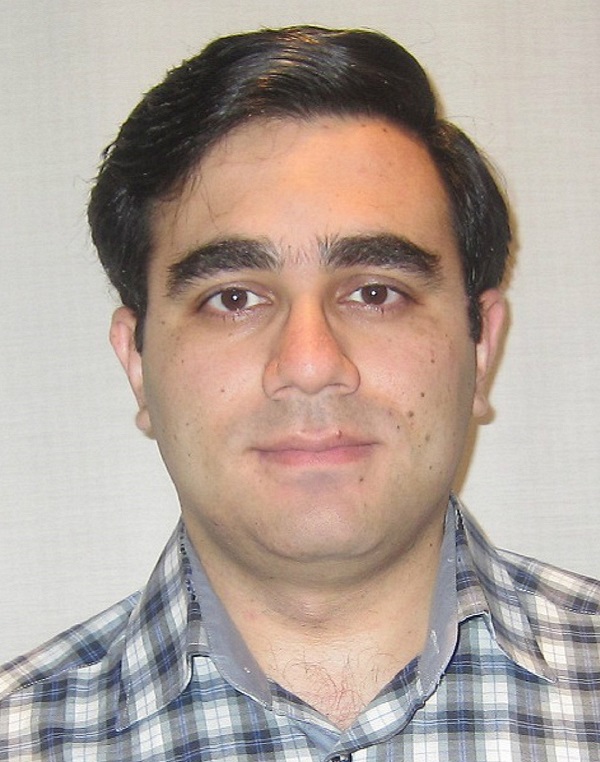}}]{Amir Alipour-Fanid} received the B.S. degree in electrical engineering from Islamic Azad University of Ardabil, Ardabil, Iran in 2005, and the M.S. degree in electrical engineering-communication from University of Tabriz, Tabriz, Iran, in 2008. 
Currently, he is working toward his Ph.D. in electrical and computer engineering department at George Mason University, Fairfax, VA, USA. 
His research interests include machine learning applications in cybersecurity, cyber-physical systems (CPS) security, Internet of Things (IoT) security, vehicle-to-vehicle (V2V) communication, and 5G wireless networks.
\end{IEEEbiography}

\begin{IEEEbiography}[{\includegraphics[width=1in,height=1.25in,clip,keepaspectratio]{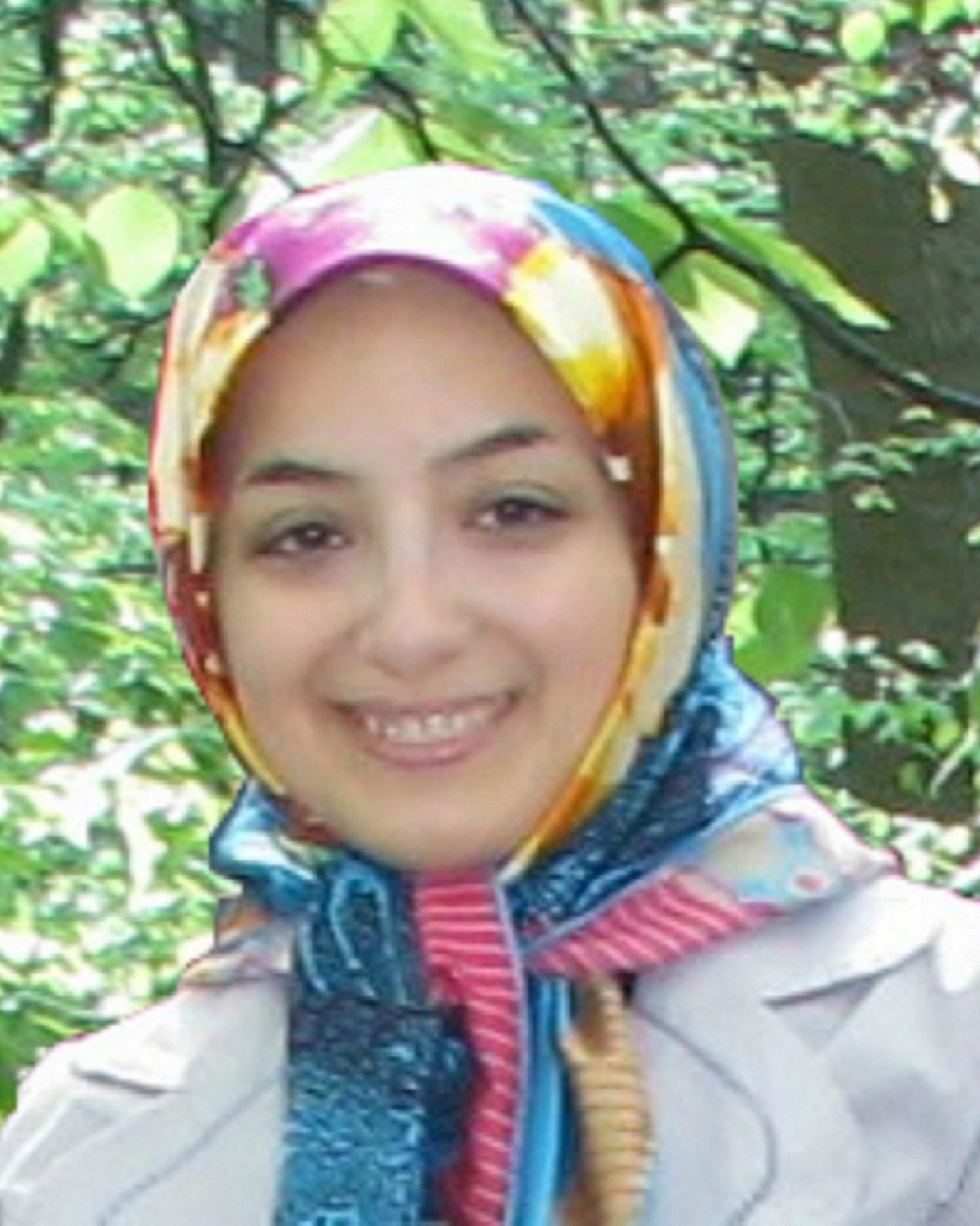}}]{Monireh Dabaghchian} received the Ph.D. degree in electrical and computer engineering from George Mason University, Fairfax, VA, in 2019. She is currently an Assistant Professor with the Department of Computer Science at Morgan State University (MSU), Baltimore, MD. She is also a member of Cybersecurity Assurance and Policy (CAP) Center at MSU. Her research interests are machine learning, deep learning, online learning, and multi-armed bandits with applications in cybersecurity, Internet of Things (IoT) security, cyber-physical systems (CPS) security, and spectrum sharing networks.
\end{IEEEbiography}

\vspace{-10mm}

\begin{IEEEbiography}[{\includegraphics[width=1in,height=1.25in,clip,keepaspectratio]{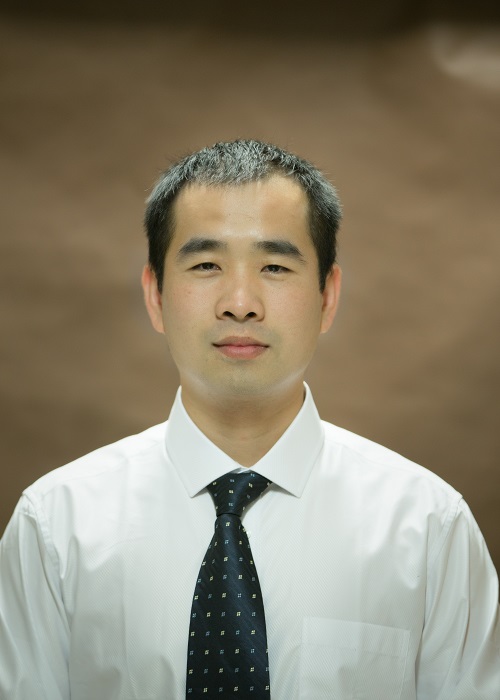}}]{Ning Wang} is currently a postdoctoral scholar in Electrical and Computer Engineering Department at George Mason University, Fairfax, VA, USA. He was an Engineer in Huaxin Post and Telecommunications Consulting Design Co., Ltd., Hangzhou, Zhejiang, Chain, from 2012 to 2013. He received the PhD degree in Information and Communication Engineering from Beijing University of Post and Telecommunication, Beijing, China, in 2017. His current research interests are in physical layer security, machine learning, device identification and RF fingerprinting.
\end{IEEEbiography}

\vspace{-10mm}

\begin{IEEEbiography}[{\includegraphics[width=1in,height=1.25in,clip,keepaspectratio]{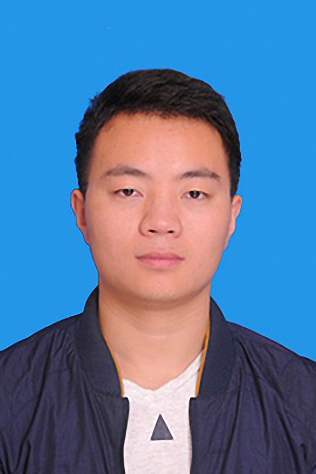}}]{Pu Wang}  received the B.S. degree in telecommunications engineering from Xidian University, Xi’an, China, in 2014, where he is currently pursuing the Ph.D. degree in cyber engineering. His current research interests include backscatter communication, wireless information and power transfer, and physical layer security.
\end{IEEEbiography}

\vspace{-10mm}

\begin{IEEEbiography}[{\includegraphics[width=1in,height=1.25in,clip,keepaspectratio]{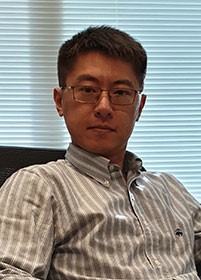}}]{Liang Zhao} is an Assistant Professor at Information Science and Technology Department of George Mason University. He received B.S. and M.S. degree from Northeastern University in China, and Ph.D. degree from Virginia Tech, USA. His research interests include spatial data mining, deep learning on graphs, sparse feature learning, interpretable machine learning, and nonconvex optimization.
\end{IEEEbiography}

\vspace{-10mm}

\begin{IEEEbiography}[{\includegraphics[width=1in,height=1.25in,clip,keepaspectratio]{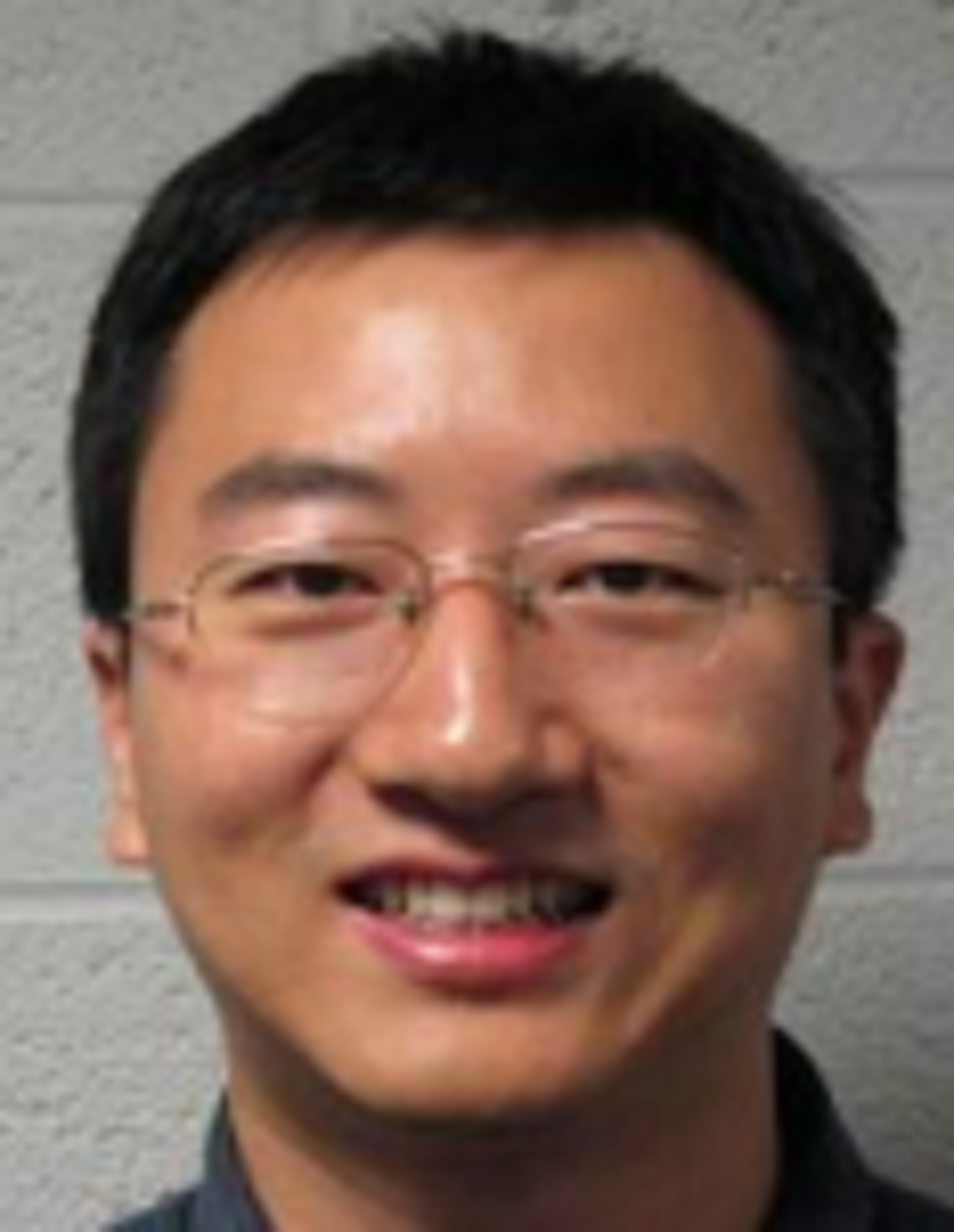}}]{Kai Zeng} received the Ph.D. degree in electrical and computer engineering from the Worcester Polytechnic Institute (WPI) in 2008. He was a Post-Doctoral Scholar with the Department of Computer Science, University of California at Davis (UCD) from 2008 to 2011. He was with the Department of Computer and Information Science, University of Michigan-Dearborn as an Assistant Professor from 2011 to 2014. He is currently an Associate Professor with the Department of Electrical and Computer Engineering, Cyber Security Engineering, and the Department of Computer Science at George Mason University. His current research interests are in cyber-physical system security and privacy, 5G physical layer security, network forensics, and spectrum sharing networks. He was a recipient of the U.S. National Science Foundation Faculty Early Career Development (CAREER) Award in 2012. He received the Excellence in Postdoctoral Research Award from UCD in 2011 and the Sigma Xi Outstanding Ph.D. Dissertation Award from WPI in 2008. He is an Editor of the IEEE Transactions on Information Forensics and Security, IEEE Transactions on Wireless Communications, and IEEE Transactions on Cognitive Communications and
Networking.
\end{IEEEbiography}

\end{document}